\documentclass[twocolumn]{aastex62}

\usepackage{amssymb}
\usepackage{amsmath}
\usepackage[]{graphicx}
\usepackage{enumerate}
\usepackage{mathrsfs,amssymb}

\usepackage{color} 
\usepackage{xspace}

\tightenlines

\begin{document}

\title{Sources and Radiations of the Fermi Bubbles}

\author{Vladimir A. Dogiel}\altaffiliation{dogiel@td.lpi.ru}
\affiliation{I. E. Tamm Theoretical Physics Division of P. N. Lebedev Institute of Physics, \\
Leninskii pr. 53, 119991 Moscow, Russia}

\author{Chung-Ming Ko}\altaffiliation{cmko@astro.ncu.edu.tw}
\affiliation{Institute of Astronomy, Department of Physics and Center for Complex Systems, \\
National Central University, Zhongli Dist., Taoyuan City 320317, Taiwan (R.O.C.)}






\begin{abstract}
Two enigmatic gamma-ray features in the Galactic central region, known as Fermi Bubbles (FBs),
were found from Fermi-LAT data. An energy release
(e.g., by tidal disruption  events in the Galactic center, GC),
generates a cavity with a shock that expands into the local ambient medium of the Galactic halo.
A decade or so ago, a phenomenological model of the FBs was suggested
as a result of routine star disruptions
by the supermassive black hole in the GC which might provide enough energy for large-scale structures, like the FBs.
In 2020, analytical and numerical models of the FBs as a process of routine tidal disruption of stars near the GC were developed,
which can provide enough cumulative energy to form and maintain large scale structures like the FBs.
The disruption events are expected to be $10^{-4}\sim 10^{-5}$ yr$^{-1}$,
providing the average power of energy release from the GC into the halo of $\dot{{\cal E}}\sim 3\times 10^{41}$ erg s$^{-1}$,
which is needed to support the FBs.
Analysis of the evolution of superbubbles in exponentially stratified disks
concluded that the FB envelope would be destroyed by the Rayleigh-Taylor (RT) instabilities at late stages.
The shell is composed of a swept-up gas of the bubble, whose thickness is much thinner in comparison to the size of the envelope.
We assume that hydrodynamic turbulence is excited in the FB envelope by the RT instability.
In this case, the universal energy spectrum of turbulence  may be developed in the inertial range of wavenumbers of fluctuations
(the Kolmogorov-Obukhov spectrum).
From our model we suppose the power of the FBs is transformed partly into the energy of hydrodynamic turbulence in the envelope.
If so, hydrodynamic turbulence may generate MHD-fluctuations,
which accelerate cosmic rays there and generate gamma-ray and radio emission from the FBs.
We hope that this model may interpret the observed nonthermal emission from the bubbles.
\end{abstract}


\section{Introduction: Sources of the Fermi Bubbles}

In this article we present our interpretation of the origin of the Fermi Bubbles (FBs).
The discussion includes the energy release in the Galactic Center (GC) to the hydrodynamic envelope in Galactic halo,
the excitation of MHD-turbulence that accelerates cosmic rays (CRs) in the halo,
the processes of nonthermal emissions which are observed in X-ray, gamma-ray and radio ranges,
and the high energy CRs escaping from the FBs to the Galactic disk.
We present a puzzle of the FB picture, where many fragments are still missing in the mosaic.
Many of them are interpreted but not completely understood yet.
The goal of the article is to find a way for a proper solution to this problem.


The origin of the energy release in the FBs in GC is still an open question.
This kiloparsec-scale structure was interpreted as a manifestation of past activity of
the central supermassive black hole (SMBH) Sgr A* in the GC.
The reader is referred to Figure 3 of \cite{predehl}, which shows the morphology of the gamma-ray bubbles (from Fermi)
and the X-ray bubbles (from eROSITA).
Observations in the GC showed structures above and below it in gamma-rays, microwave and X-rays.
The eROSITA \citep[][]{predehl} found giant bubbles in the X-ray range $0.1\sim 2.4$ keV
extending approximately 14 kiloparsecs above and below the GC.
The estimated energy of the bubbles is around $10^{56}$ erg.
The total luminosity in X-rays is about $10^{39}$ erg s$^{-1}$ which could be the result of past activities in the GC.
The temperature of the envelope is about $0.3$ keV,
the velocity of the shock is about $340$ km s$^{-1}$ or of Mach number $\approx 1.5$,
and the energy-release rate of the gas envelope is roughly $10^{41}\sim 10^{42}$ erg s$^{-1}$.

The inner radius of the X-rays shell (about 7 kpc) coincides spatially with the region of GeV gamma-rays
in the range $1\sim 100$ GeV with the luminosity of $F_\gamma \approx 4\times 10^{37}$ erg $s^{-1}$ \citep[][]{su10}.
The bubble structure in the GC was also revealed in the range of microwaves which coincides nicely with that of gamma-rays \citep[][]{planck13}.
The flux is in the range $23\sim 61$ GHz, and the luminosity is $\Phi_\nu\approx 1\sim 5\times 10^{36}$ erg s$^{-1}$.


Similar giant structures near the GC were found earlier in the radio in hundred MHz \citep[the North Polar Spur, see][]{sof77}
and in 1.5 keV X-ray emission \citep[see][]{bland03}.
These structures were postulated as bipolar supershells which were produced by starbursts.
A shock front was supposed to reach a radius 10 kpc in the polar regions which could be consistent with the GC explosions.
This model required an energy release of about $10^{55}$ erg at the GC, and periodic activity on a timescale of $10\sim 15$ Myr.

\citet{mou23} suggested that the nature of North Polar Spur in the GC agreed with the eROSITA bubble of ages about 20 Myr \citep[][]{predehl}.

\begin{figure}[ht]
\centering
\plotone{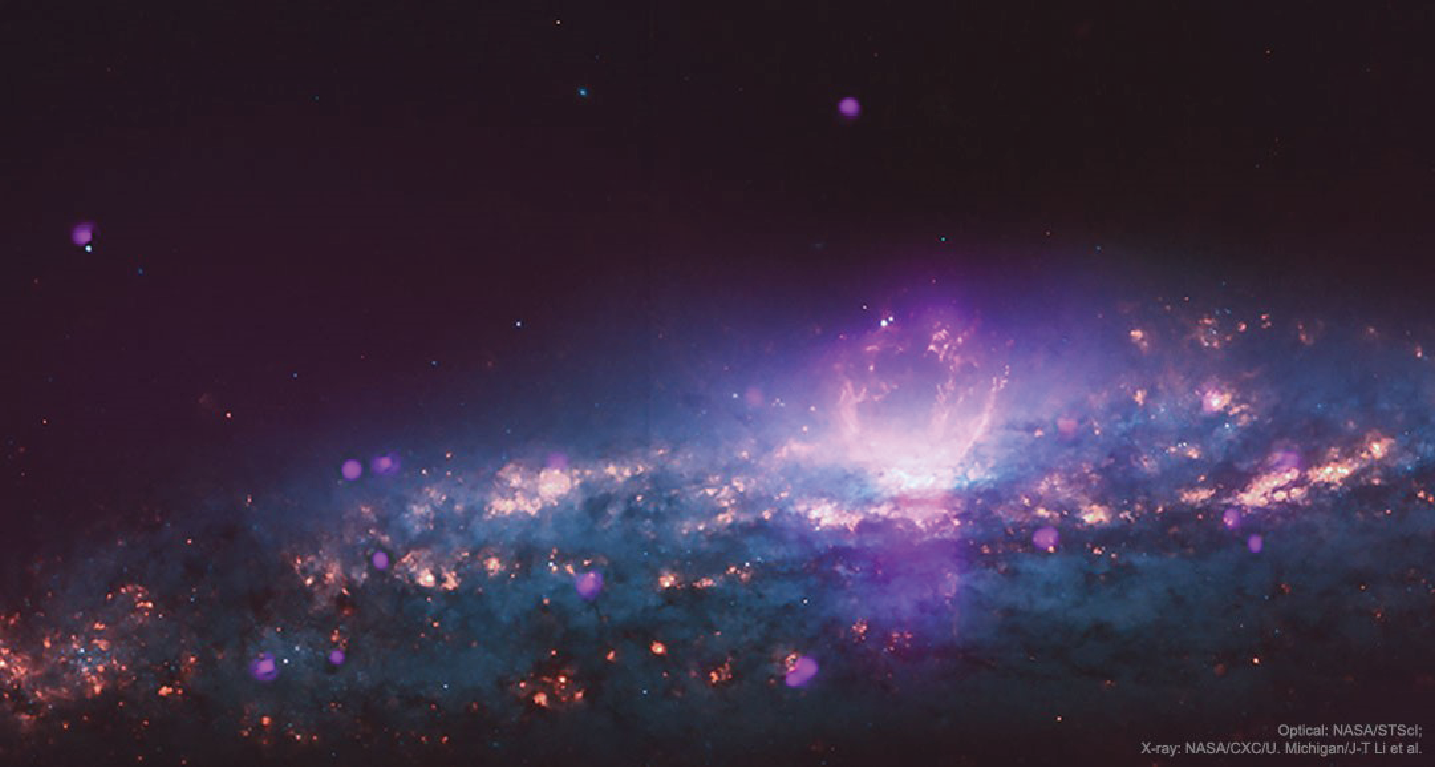}
\caption{
Two superbubbles in the galaxy NGC 3079 observed in X-ray (purple and pink).
Image obtained from
{\tt
https://chandra.harvard.edu/photo/2019/ngc3079
}
\!\!\!\!.
Image credit: X-ray: NASA/CXC/University of Michigan/J-T Li et al.; Optical: NASA/STScI.
}
\label{fig:NGC3079}
\end{figure}

Bubbles were also discovered in other galaxies \citep[see review by][]{sarkar2}.
Fig. \ref{fig:NGC3079} shows superbubbles from the galaxy NGC 3079.

The total energy needed to generate large Galactic outflows is assumed to be in the range up to about $10^{56}$ erg.
This energy release in the GC may be compelling evidence for a huge energetic explosion occurred in the GC a few ($2\sim 8$) million years ago,
\citep[see, e.g.,][and references therein]{bland03}.
For examples, \citet{nayak18} assumed a capture of a giant molecular cloud of mass $\sim 10^5M_\odot$ in the GC about one Myr ago;
and \citet{yang2012,yang22} suggested a model of FBs as a results of the past activity at the GC.

Alternative models of the bubbles were suggested by, e.g., \citet{cheng06,cheng07,cheng11}, \citet{sarkar}, \citet{zub12}, \citet{ko20}, etc.
They suggested that the source of energy of the bubbles is sporadic energy releases in the GC
by stellar tidal disruption events (TDEs) near the central SMBH
\citep[see also other suggestions such as active star-formation near the GC, e.g.,][]{zhang24}.
The motion of nearby stars orbiting around Sgr A* (the GC) has been observed more than two decades \citep[][]{ghez2005,gil09,genz10}.
Analysis of the motions gave an
estimated mass of about $4.4 \times 10^6\ M_{\odot}$ for the central SMBH.
For an illustration of orbits of stars around Sgr A*, see Figure 16 of \citet{gil09}.
For the motion of stars orbiting Sgr A*, the reader is referred to
{\tt
https://www.eso.org/public/videos/eso1825e
}
(or
{\tt
https://www.youtube.com/watch?v=TF8THY5spmo
}
);
and for animation of stellar orbits around Sgr A* to
{\tt
https://www.eso.org/public/videos/eso1825f
}
(or
{\tt
https://www.youtube.com/watch?v=wyuj7-XE8RE
}
),
{\tt
https://www.youtube.com/watch?v=tMax0KgyZZU
}
\!\!\!\!.


TDE occurs when a star is getting too close to a SMBH (closer than the tidal radius).
The classic picture is the star is disrupted by the tidal force, and after half of the stellar debris is in unbound orbits
and the other half in bound orbits and fallback towards black hole \citep[][]{Rees1988}.
The problem is how and how much energy is released into the host galaxy.
For instance, how much energy is carried away by the unbound debris,
and how much binding energy is released by the bound debris, say through accretion.
Theoretically, it is possible to have $5\%$ of the rest mass energy of the star being released
(for a solar type star it is about $10^{53}$ erg).
However, the results from observations are mixed, from several $10^{51}$ to $10^{53}$ erg,
It is an active area of research to study TDE from different perspectives
\citep[to name a few, e.g.,][]{Burrows2011,Zauderer2011,donato14,piran,Metzger2016,kara16,lin17,dai18,Lu2018,Mockler2021,Goodwin2023}.
The reader is also referred to reviews like \citet{dai21} and \citet{Gezari2021}.
Another issue is the rate of TDEs at the centre of a galaxy with a SMBH.
It is conceivable that the rate depends on the type of galaxy and the environment near the black hole.
A typical estimation is roughly $10^{-4}\sim 10^{-5}$ yr$^{-1}$ \citep[e.g.,][]{stone16,Hung2018,sarkar1}.

\citet{ko20} adopted the TDE model of \citet{dai18} and obtained the outflow energy of an event is about $10^{52}\sim 10^{53}$ erg
(the outflow velocity is about $0.1\sim 0.3$ the speed of light).
Together with an event rate about $10^{-4}$ yr$^{-1}$ will provide an average power about $\dot{\cal{E}}\sim 3\times 10^{41}$ erg s$^{-1}$,
which will be sufficient to power the FBs.
\citet{mill16} inferred a bubble expansion rate of 490 km s$^{-1}$, an age of 4.3 Myr,
and a luminosity $2.3\times 10^{42}$ erg s$^{-1}$ \citep[see also the review of FBs in][]{sarkar2}.


Recently, evidence of the energy release at Sgr A* was interpreted as the result of the latest stellar disruption.
Two elongated chimneys of about 150 pc near the GC were found in the X-ray \citep[][]{ponti19,ponti21} and radio \citep[][]{hey19} ranges.
In both cases the total energy of the chimneys was estimated to be
about $10^{53}$ erg or below,
which might be a result of the latest TDE by the central SMBH.

\begin{figure}[ht]
\centering
\plotone{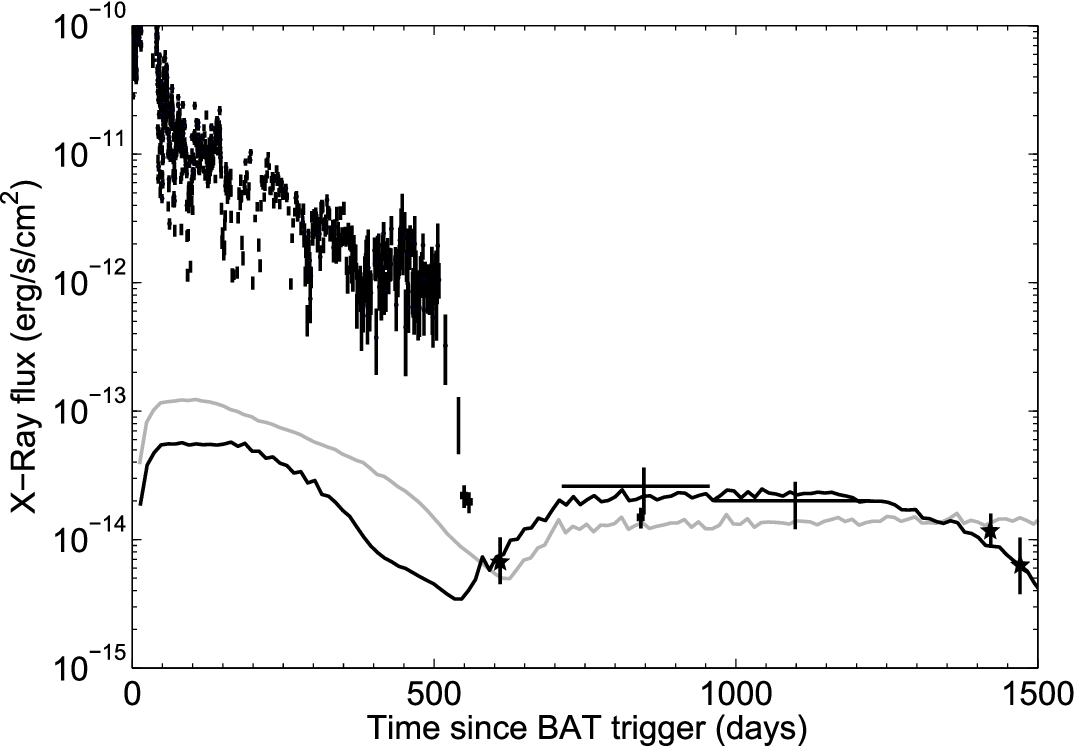}
\caption{
The observed X-ray light curve of Swift J1644+57
from Swift, XMM-Newton and Chandra.
Figure reproduced from \citet{cheng16} with permission.
}
\label{swift}
\end{figure}

Similar processes of stellar disruption at the galactic center of other galaxies were observed.
For example, X-ray transient Swift J164449.3+573451 (also known as GRB 110328A) was detected by Swift in the direction of the constellation Draco
with the peak luminosity $10^{48}$ erg s$^{-1}$ (see Fig. \ref{swift}).
Observations showed that the transient originated from the center of a galaxy at cosmological distances involving a SMBH in the galaxy nucleus.
It was concluded that Swift J164449.3+573451 is most likely originated from the central SMBH,
and the X-ray and radio emissions were interpreted as a result of stellar capture by the black hole
\citep[see, e.g.,][]{levan,Burrows2011,Zauderer2011}.

\section{Structure of the Fermi Bubbles}
\label{sec:structure}

A sudden (sporadic) energy release by a TDE in the GC creates a cavity with a shock
which expands into the surrounding non-uniform medium of the halo.
For example, the gas distribution in the halo above (and below) the Galactic plane is decaying exponentially with a scale height
$H=2$ kpc and the density at the plane is $n_0=4\times 10^{-3}$ cm$^{-3}$ \citep[see, e.g.,][]{Nakashima2019}.
The exponential gas distribution is
\begin{equation}
{\cal R}(z)=\exp\left(-\frac{z}{H}\right)\,,
\label{eq:exp}
\end{equation}
where ${\cal R}(z)=n(z)/n_0$.

Alternatively, \citet{mill16} suggested a so-called $\beta$-model of the gas density profile in the halo from the intensity of absorption lines,
\begin{equation}
{\cal R}(z)= \left(\frac{z}{z_c}\right)^{-3\beta}\,,
\end{equation}
where $z_c=0.26$ kpc and $n_0=0.5$ cm$^{-3}$.

\begin{figure}[ht]
\centering
\plotone{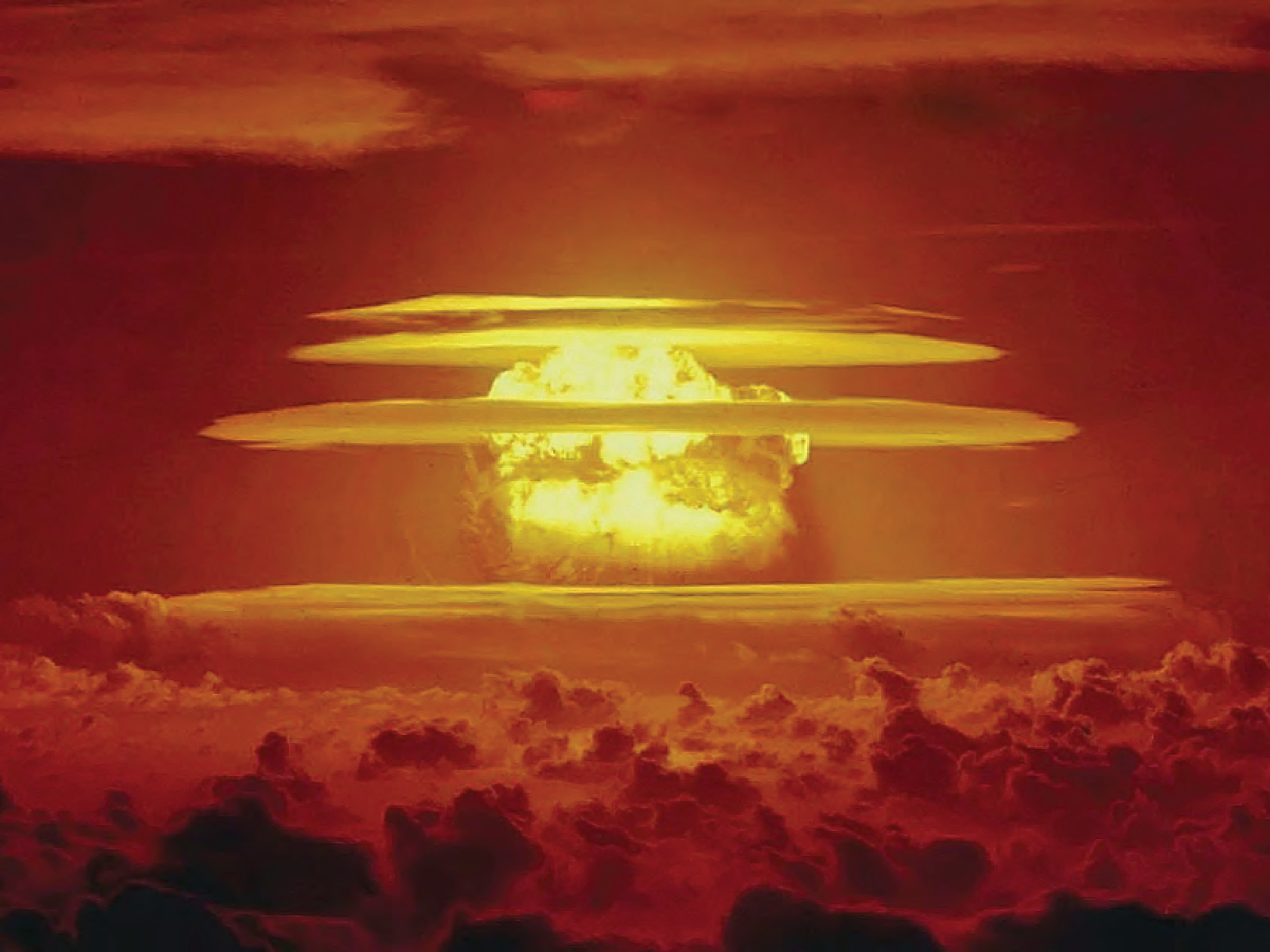}
\caption{
A thermonuclear explosion in the terrestrial atmosphere.
Image credit: United States Department of Energy.
Image from
{\tt
https://commons.wikimedia.org/wiki/
File:Castle$\_$Bravo$\_$nuclear$\_$test$\_$(cropped).jpg
}.
}
\label{hbom}
\end{figure}

The formalism of envelope propagation was developed as a solution for a strong explosion \citep[see][]{sed59},
and was elaborated by \citet{komp60} for explosion in a non-uniform atmosphere.
The reader is referred to the review of \citet{kogan} and the monograph of \citet{zeld67}.
This model was derived for the thermonuclear explosion in the terrestrial atmosphere.
Fig. \ref{hbom} shows an example of a thermonuclear explosion test in the terrestrial atmosphere.

\begin{figure}[ht]
\centering
\plotone{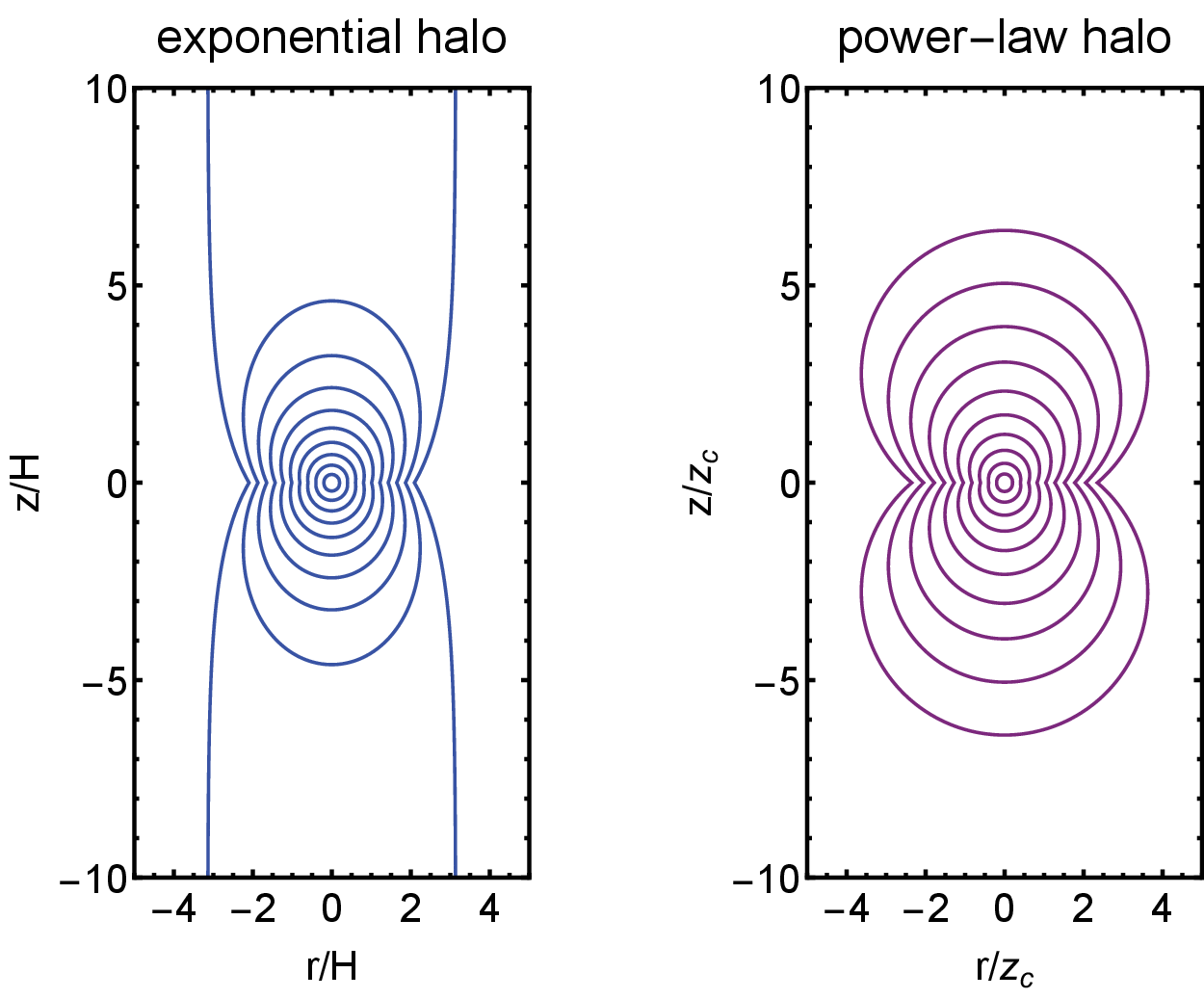}
\caption{
Illustration of the double-bubble shock envelope in the halo evolving with time.
The gas distribution in the halo in the left panel is exponential and in the right panel is power-law.
Figure adapted from \citet{ko20} with permission.
}
\label{fig:evolve}
\end{figure}

\citet{fk98}, \citet{baumbr13}, \citet{ko20} and \citet{breit22}
developed analytical solutions of a hydrodynamic model for the shock wave propagation
in non-uniform atmospheres or halos, for different energy input rates for single and successive explosions.
The shock envelope generated has a double-bubble structure in the halo (see Fig. \ref{fig:evolve}).

Following the Kompaneets formalism \citep[see details in][]{kogan}, the shock front is described as
\begin{equation}
\left(\frac{\partial r}{\partial y}\right)^2
-\frac{1}{{\cal R}(z)}\left[\left(\frac{\partial r}{\partial z}\right)^2+1\right]=0\,,
\end{equation}
where $y(t)$ is a transformed time (in units of length)
\begin{equation}
y=\int_0^t\sqrt{\frac{\left(\gamma_{\rm g}^2-1\right)}{2}\frac{2{\cal E}(t)}{3\rho_0V(t)}\,}\,dt\,,
\end{equation}
$V(t)$ is the bubble volume
\begin{equation}
V(t)=\pi\int_0^{z_u}r^2(z,t)dz\,,
\end{equation}
with $r(z,t)$ or $r(z,y)$ as the bubble radius at the altitude $z$,
$\rho_0=n_0 m_p$ is the mass density corresponding to $n_0$
(the number density at the base $z=0$),
${\cal E}$ is the energy released by the central source into the bubble,
and $\gamma_{\rm g}$ is the adiabatic index of the gas.

For the exponential gas distribution given by Eq. (\ref{eq:exp}), the top of the bubble $z_u$ is a function of time $t$,
\begin{equation}
z_u=-2H\,\ln\left(1-\frac{y}{2H}\right)\,.
\end{equation}
In this model of \citet{baumbr13} and \citet{breit22} the velocity at the top of the bubble $v_u$ for  the total energy release of
${\cal E}$ is
\begin{equation}
v_u=\frac{d z_u}{d t}=\exp\left(\frac{z_u}{2H}\right)\frac{d y}{d t}\,.
\end{equation}

In the early phase when $z_u \lesssim H$, the expanding cavity can be described by the Sedov solution \citep{sed59}
when the gas density is almost uniform and the velocity of the shock envelope decreases with time.
When $z_u > H$ the shock propagates in the exponential halo with acceleration afterwards \citep[see][]{baumbr13}.

The propagation of the shock envelope is derived under the strong shock assumption.
In reality, if the velocity of the envelope is below the sound speed $c_s$ of the halo gas,
then the shock or the envelope will decay and be absorbed in the halo.
On the other hand, if this velocity is higher than the sound speed, the shock will be able to penetrate into the halo,
and transfer the energy from the initial central source into the exponential halo.
The velocity of the top of the bubble is the fastest,
\citet{baumbr13} defined a condition of shock penetration into the exponential halo: $v_u(y_{\rm acc})>3c_s$,
where $v_u(y_{\rm acc})$ is the minimum of $v_u$ and this occurs at $y=y_{\rm acc}$,
i.e., ${\dot v}_u(y_{\rm acc})={\ddot z}_u(y_{\rm acc}) = 0$.
The acceleration at the top of the bubble is \citep[see][]{baumbr13},
\begin{equation}
\ddot{z}_u=\frac{H}{t^2_{\rm SN}}\,\frac{\left(\gamma_{\rm g}^2-1\right)}{4(1-\tilde{y}/2)\tilde{V}}
\left[\frac{1}{\left(1-\tilde{y}/2\right)}-\frac{1}{\tilde{V}}\frac{d\tilde{V}}{d\tilde{y}}\right]\,,
\label{accel1}
\end{equation}
where
\begin{equation}
\tilde{y}=\frac{y}{H}\,,\quad
\tilde{V}=\frac{V}{H^3}\,,\quad
t_{\rm SN}=\sqrt{\frac{3\rho_0 H^5}{2{\cal E}_{\rm SN}}}\,.
\end{equation}

\begin{figure*}[ht]
\centering
\plotone{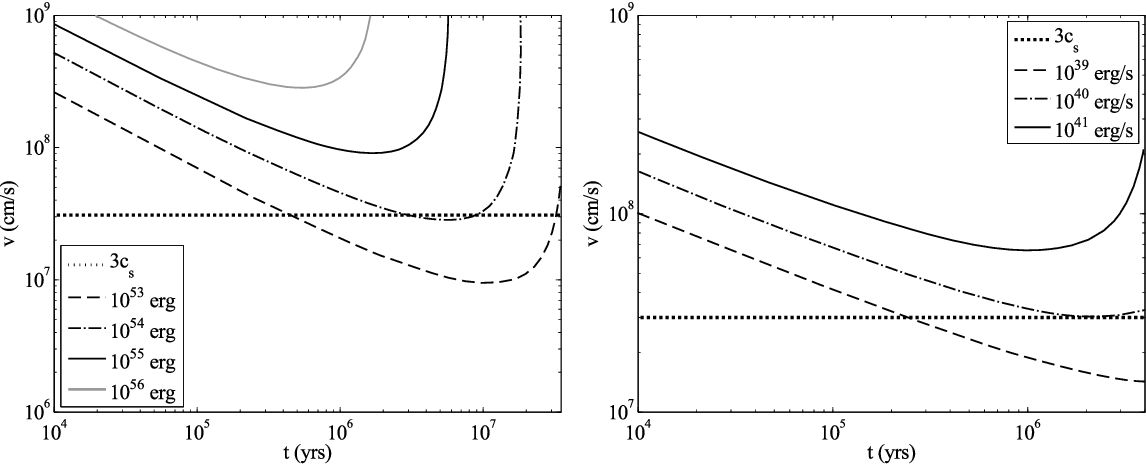}
\caption{
Temporal variation of the shock velocity of the top of the bubble for the case of exponential halo with $H=0.67$ kpc and $n_0=0.03$ cm$^{-3}$.
{\it Left panel}: One single input of energy from the GC.
{\it Right panel}: Multiple TDEs with different values of the power release at the GC.
The horizontal dotted line indicates the velocity which is necessary for the shock in order not to stall in the halo.
It is three times the sound speed in the halo $3\times 10^7$ cm s$^{-1}$.
Figure reproduced from \citet{ko20} with permission.
}
\label{veloc1}
\end{figure*}

As an example we present in the left panel of Fig, \ref{veloc1}, the development of the shock velocity at the top of the bubble
from a single energy explosion ${\cal E}$.
The energy of the explosion occupies more and more volume of the exponential atmosphere, and finally approaching infinity in finite time
(provided that the velocity is always larger than $3 c_s$).

For the parameters in the GC, a single star disruption event provides no more than $10^{52}\sim 10^{53}$ erg \citep[see][]{piran,dai18,dai21}.
There is not enough energy for the FBs or similar structures.
An unusually huge single energy release in the past, say exceeding ${\cal E}>10^{54}$ erg, may explain the origin of the Fermi Bubbles.

Alternatively, this huge energy can be supplied by a series of many weaker disruption events with an effective power input
$\dot{\cal E}\geq 10^{40}$ erg s$^{-1}$, see the right panel of Fig. \ref{veloc1}.
It may be interpreted as routine TDEs of which each produced an energy of $10^{52}\sim 10^{53}$ erg,
and the average rate of stellar capture is about $10^{-5}\sim 10^{-4}$ yr$^{-1}$ \citep[see][]{ko20}.

The envelope shell is composed of the swept-up gas of the bubble, and it is much thinner in comparison to the size of the bubble.
The shell thickness is defined as
\begin{equation}
d(y)=\frac{M_s(y)}{2\pi\rho_{{\rm sh}0}\int_0^{z_u(y)}e^{-z/H}r(z,y)\sqrt{1+\left(\frac{\partial r}{\partial z}\right)^2\,}dz}\,,
\end{equation}
where $M_s$ is the total mass of the FB
\begin{equation}
M_s(y)=\pi \rho_0\int_0^{z_u(y)}e^{-z/H}r^2(z,y)dz\,,
\end{equation}
and $\rho_{\rm sh}=\rho_{{\rm sh}0}\,\exp(-z/H)$ is the density within the shell \citep[see][]{breit22}.

\begin{figure}[ht]
\centering
\plotone{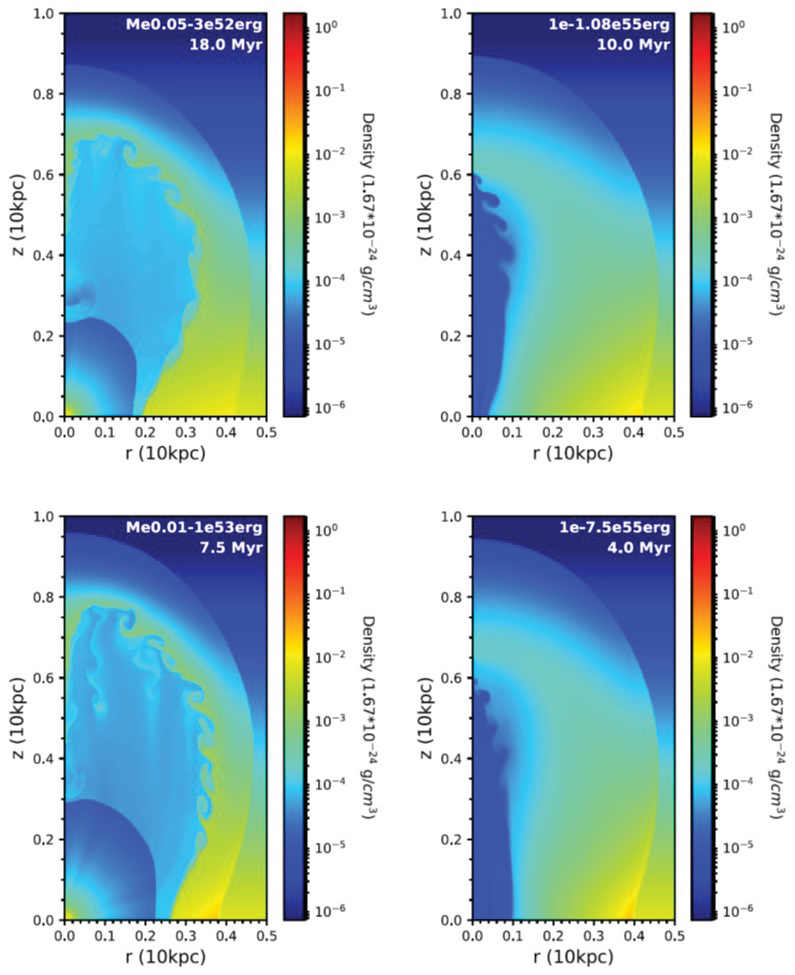}
\caption{
Density distribution of numerical simulations of the FBs in an exponential halo.
The two panels on the left column are results of multiple explosions (e.g., TDEs) and the right column are results of a single huge explosion.
In the upper left panel, ``Me0.05-3e52erg 18.0 Myr'' corresponds to multiple explosions with 0.05 Myr between successive explosions
and the energy release by each explosion is $3\times 10^{52}$ erg, and the simulation ends at 18.0 Myr.
In the upper right panel, ``1e-1.08e55erg 10.0 Myr'' corresponds to a single explosion with an energy release of $1.08\times 10^{55}$ erg,
and the simulation ends at 10.0 Myr.
Similar explanation for the lower panels.
Lower panel figures reproduced from \citet{ko20} with permission.
}
\label{gas}
\end{figure}

Fig. \ref{gas} shows some examples of numerical simulation of the FB envelope from sporadic star disruption or from a single huge explosion.

\section{Envelope Disruption by Rayleigh-Taylor Instability}
\label{sec:RT_instability}

The interface between a denser fluid supported by a lighter fluid in a gravitational field is susceptible to Rayleigh-Taylor (RT) instability.
The amplitude of an infinitesimal perturbation will grow exponentially at the early phase or the linear phase \citep[see, e.g.,][]{chandra}.
The growth time of the instability in linear phase is
\begin{equation}
\tau_{\rm RT}=\sqrt{\frac{\lambda}{2\pi g}\,\frac{(\rho_2+\rho_1)}{(\rho_2-\rho_1)}\,}\,,
\end{equation}
where $\lambda$ is the wavelength of the perturbation, $g$ is the gravitational acceleration
and $\rho_1$ and $\rho_2$ are the densities of the lighter and denser fluids, respectively.
For an illustration of time evolution of the RT instability, the reader is referred to Figure 4 of \citet{breit22}.


In the case of a superbubble,
the RT instabilities are excited between the dense shell and the hot interior
when the envelope is accelerating into the exponential halo \citep[see][and Eq. (\ref{accel1})]{baumbr13,breit22}.
Identifying the gravitational acceleration with the acceleration at the top of the bubble ${\ddot z}_u(y)$
gives the growth time of the instability (in this coordinates)
\begin{equation}
\tau_{{\rm RT},z_u}({y})=\sqrt{\frac{d(y)}{2\pi {\ddot z}_u({y})}
\frac{[\rho_{\rm sh}(y)+\rho_{\rm in}(y)]}{[\rho_{\rm sh}(y)-\rho_{\rm in}(y)]}\,}\,,
\end{equation}
where $y$ is the transformed time (see Section \ref{sec:structure}).
When the wavelength $\lambda$ of the RT fluctuations is about the envelope shell thickness $d$,
these instabilities may destroy the bubble,
see Figure 13 of \citet{breit22}.

The temporal evolution of the RT instability during the nonlinear regime is obtained by
numerically solving the ordinary differential equation for the RT fluctuations of $\lambda$,
\begin{equation}
{\dot\lambda}(y)=2\sqrt{\alpha {\ddot z}_u(y)\lambda(y)\,}\,.
\label{zdot}
\end{equation}
The parameter $\alpha$ is estimated from the initial condition
$\lambda(y_0)\sim 0.01 d(y_0)$ \citep[see][]{baumbr13}.


\section{Energy and Spectrum of Hydrodynamic Fluctuations}
\label{sec:hydrodynamic_turbulence}

From Eq. (\ref{zdot}) we can estimate the fraction of the total energy of the FBs,
$\dot{\cal E}\sim 3\times10^{41}$ erg s$^{-1}$,
that is transformed into the hydrodynamic turbulence in the envelope excited by the RT instabilities there.
From \citet{land} we get the rate of energy dissipation in the turbulent of flux
\begin{equation}
\varepsilon(t)=\frac{v_\lambda^3}{\lambda}\,.
\end{equation}
Instead of $\lambda$ we can introduce a wavenumber $k=2\pi/\lambda$.
(the Kolmogorov-Obukhov spectrum of turbulence).
Then the kinetic energy spectrum $W(k)$ of the turbulence is
\begin{equation}
W(k)\sim \varepsilon^{2/3}k^{-5/3}\,,
\label{5-3}
\end{equation}
and
\begin{equation}
\int_k^\infty W(k) dk\sim v_\lambda^2\,.
\label{eq:int_W}
\end{equation}
In this case a universal energy spectrum $W(k)$ is developed in the inertial range as shown in Eq. (\ref{5-3}),
i.e., the Kolmogorov-Obukhov spectrum of turbulence \citet{land}.

The energy losses of RT, $\varepsilon(t)$, in the envelope, transferred into the turbulence, is
\begin{equation}
\varepsilon(t)=\frac{{\dot\lambda}^3_0(t)}{\lambda_0(t)}=8\alpha^{3/2}{\ddot z}_u^{3/2}\lambda^{1/2}_0\,,
\label{lum}
\end{equation}
where $\lambda_0 =2\pi/k_0$ is the pumping scale
(see below).


The spectrum of hydrodynamic fluctuations $W(k)$ has a wide inertial interval,
where the energy is supplied in the initial scale of $k_0<k$, in which only energy transfer along the spectrum is realized.
In the energy range the spectrum at
$\lambda<\lambda_0=2\pi/k_0$
is determined only by
the power of energy pumping at the scale $\lambda_0$.

The growth of this large-scale structure may be understood at each stage in terms of ring vortex pairing, mutual orbiting,
and merging, followed by these processes repeating with the just merged eddies on a larger scale.
During this development, ambient material is entrained and intense smaller scale turbulence is generated in the regions between the vortices,
presumably establishing the turbulent cascade to higher wavenumbers, which is eventually dominated by viscosity on the Kolmogorov microscale, $l_K$.
The small-scale of the stretched turbulence-generating regions between the vortices can be associated with the Taylor microscale, $l_T$.

The total fluid turbulence energy is given by $E_t=\rho v^2_t$, where $\rho$ is the fluid density
and $v_t$ is the root mean square of the velocity of the turbulence (or simply the turbulence velocity)
at the largest scale $\lambda_0$ or $l_0$.
The range is taken to extend from $l_0$, down to the effective damping (or Kolmogorov) scale $l_K$.
The Reynolds number in this case is,
\begin{equation}
Re=\left(\frac{l_0}{l_K}\right)^{4/3}\,,
\label{eq:Reynolds}
\end{equation}
where the smallest scale, $l_K$, is defined by the dissipation of turbulence.

There is no absolute definition of the scale $l_T$ of this transition
from hydrodynamic spectrum to that of MHD (upper limit of MHD-turbulence scale).
Following \citet{eilek84} the Taylor length, $l_T$, was estimated as
\begin{equation}
l_T\sim l_0\left(\frac{15}{Re}\right)^{1/2}\,.
\end{equation}
Therefore, the fluid cascade extends from the Taylor scale $l_T$, on which the transition from a large-scale ordered turbulence to
smaller scale disordered motion occurs. The cascade proceeds down to the smallest scale $l_K$, determined by dissipation,
\begin{equation}
l_K \approx l_0\,(Re)^{-3/4}\,.
\end{equation}

We adopt the hydrodynamic view that the Kolmogorov equilibrium cascade exists between $l_T$ and $l_K$ as our first approximation
to the complex interactions expected in the regime.
The relation between (the root mean square of) the turbulence velocity $v(l)$ and the turbulence scale $l$ is
\begin{equation}
v(l)=v(l_T)\left(\frac{l}{l_T}\right)^{1/3}\,.
\end{equation}

\section{Particle Acceleration by Alfv\'en Fluctuations and the Lighthill Radiation}
\label{sec:Lighthill_radiation}

Contrary to shock acceleration, the turbulent resonant acceleration does not require strong shocks.
The fluid turbulence in the interstellar medium or intercluster medium is another possible origin of particle acceleration,
e.g., in galaxy clusters \citep{brun04}, and in radio jets in turbulent mixing regions \citep{hendr}.
Turbulent motions act as a source of waves which is presented as a hierarchy of eddies.
The Lighthill mechanism acts as a direct source of energy to the MHD waves over the range of wavenumbers
corresponding to the fluid turbulent spectrum.
A turbulent eddy has kinetic energy, which it releases when it mixes with its surroundings.
Most of this energy is returned to the ambient medium, but a small fraction gets transformed
into propagating waves \citep[see, e.g.,][and others]{stein,hendr,blasi00,brun,fuji}.
A strong coupling between particle energy and turbulent energy spectra can be expected,
and the hydrodynamic turbulence in the medium accelerates particles through wave-particle resonance.
Alfv\'en waves is an alternative source of charged particle acceleration
via resonant interaction of MHD-waves and with relativistic particles.

In the pioneering paper \citet{light} developed the model of acoustic waves
which are excited by hydrodynamic turbulence in the absence of magnetic fields.
This is known as the Lighthill radiation.
The radiated power of the waves is roughly the energy density in the turbulent motion, $\varepsilon$,
divided by the decay time scale of waves $\tau$,
and the compactness of the eddy (which is measured by the ratio of the size of the eddy, $l$, to the wavelength, $\lambda=2\pi/k$).

\citet{kuls55} showed that this radiation of MHD-turbulence is excited if there is an external constant magnetic field.
If there is no external magnetic field, the magnetic turbulence generates sound waves only via the Lighthill mechanism.
If there is a constant external magnetic field, hydromagnetic waves are generated instead by Alfv\'en waves
unless the energy density of hydrodynamic turbulence prevails over the energy of magnetic density.
The central idea was a coupling between the hydrodynamic eddy cascade and the MHD waves through the process of Lighthill radiation
is presented in \citet{kuls55}, \citet{parker} and \citet{kato}.

\citet{kato} developed the Lighthill theory of the MHD-radiation for strong and weak magnetic fields $B$
which is characterized by the magnetic Mach number,
the ratio of the (root mean square) velocity of the hydrodynamic turbulence $v$ to the Alfv\'en speed $v_A=B/\sqrt{4\pi \rho}$,
\begin{equation}
M_A=\frac{v}{v_A}\,.
\label{eq:MachA}
\end{equation}

For small Mach numbers ($M_A\ll 1$)
a small fraction of the power emitted is in the form of Alfv\'en waves,
while for large Mach numbers ($M_A\gg 1$)
the power emits as sound waves and the radiation of Alfv\'en waves is insignificant.
The turbulence decay time scale is the nonlinear cascade time, which is the eddy turnover time,
i.e., the eddy size $l$ divided by its velocity
$v$,
\begin{equation}
\tau\approx \frac{l}{v}\,.
\end{equation}

For Alfv\'en waves in strong magnetic field, and the frequency is $\omega=\bar{k}v_A$, and
\begin{equation}
\bar{k} l\approx \frac{v}{v_A}\,,
\end{equation}
which corresponds to the resonance $\tau(l)\approx 1/\omega(\bar{k})$.

From radio polarization measurements, \citet{zhang24} showed that there are large-scale magnetic fields in the Fermi and eROSITA bubbles.
Figure 1 of \citet{zhang24} showed several kpc-scale magnetised structures in the bubbles.


If the turbulent magnetic field dominates the motion ($M_A\ll 1$), then the power is
\begin{equation}
P_{A}\sim \frac{\rho v^3}{l}M_A\,,
\label{eq:PA}
\end{equation}
and a small fraction of flux of hydrodynamic is transformed into Alfv\'en waves.
In the following we present some details of the radiation of MHD-waves
in space medium in the limit of strong external magnetic fields.

We assume that fluid turbulence is induced by the
motion of a smaller cluster in a larger cluster and its energy
spectrum is described by a power law \citep[see][]{eilek84,fuji}
\begin{equation}
W_f(k)=W_f^0k^{-m}\,,
\label{eq:turbulence}
\end{equation}
where $k=2\pi/l$ is the wavenumber corresponding to the scale $l$,
$W_f(k)\delta k$
is the energy per unit volume in turbulence with wavenumbers between $k$ and $k+\delta k$,
and $W_f^0$ and $m$ are the constants.
If one expresses the turbulent spectrum in terms of eddy size, the spectrum is represented by $W(l)\propto l^{m-2}$.
The cascade of the fluid turbulence extends from a largest eddy size $l_0=2\pi/k_0$ down to a smallest scale determined by dissipation
$l_K\sim l_0Re^{-3/4}$ (where $Re$ is the Reynolds number).
Since most of the energy of fluid turbulence resides in the largest scale,
the total energy density of fluid turbulence is presented in the form as $E_t~\sim \rho v_t^2$, where $\rho$ is the fluid
density and $v_t$ is the turbulent velocity of the largest scale $l_0$.
The normalization $W_f^0$ can be derived from the relation
$E_t =\int_{k_0}^{k_T} W_f(k)dk$,
\begin{equation}
W^0_f= \frac{E_t}{R}k_T^{(m-1)}\,,
\end{equation}
where
\begin{eqnarray}
& R &=\frac{1}{(m-1)}\left[\frac{k_0W_f(k_0)}{k_TW_f(k_T)}-1\right] \nonumber\\
& & \approx \frac{1}{(m-1)}\frac{k_0W_f(k_0)}{k_TW_f(k_T)}\,.
\end{eqnarray}
Here $k_T=2\pi/l_T$ and $l_T$ is the wavelength below which Alfv\'en waves are driven.

A fluid eddy of size $l$ has a velocity
\begin{eqnarray}
& v(l) &\approx \left[\frac{lW_f(l)}{\rho}\right]^{1/2}=\left[\frac{kW_f(k)}{\rho}\right]^{1/2} \nonumber\\
& & =\left(\frac{E_t}{\rho R}\right)^{1/2}\left(\frac{k}{k_T}\right)^{(1-m)/2}\,.
\label{vf}
\end{eqnarray}
Turbulence on a scale $k$ will radiate Alfv\'en waves at the wavenumber
\begin{equation}
\bar{k}=\left[\frac{v(l)}{v_A}\right]k\,,
\label{6}
\end{equation}
Here we recall that $k$ is the wavenumber of hydrodynamic turbulence and $\bar k$ is the wavenumber of the Alfv\'en waves.

Let $v[k(\bar{k})]$
be the fluid velocity on the fluid scale $k(\bar{k})$ that drives Alfv\'en waves of wavenumber $\bar{k}$.
From equations (\ref{vf}) and (\ref{6}), we obtain
\begin{equation}
v[k(\bar{k})]=v_A\left[\frac{E_t}{\rho v_A^2R}\left(\frac{\bar{k}}{k_T}\right)^{(1-m)}\right]^{1/(3-m)}\,.
\label{7}
\end{equation}

Assuming that the energy going into Alfv\'en waves at wavenumber $\bar{k}$ with an energy flux
\begin{equation}
I_A(\bar{k})=I_0(\bar{k}/k_T)^{-s_t}\,,
\label{ia}
\end{equation}
where $I_A(\bar{k})\delta\bar{k}$ is the power per unit volume going into Alfv\'en waves with wavenumbers
in the range $\bar{k}\rightarrow \bar{k}+\delta\bar{k}$ and $I_0$ and $s_t$ are the constants.
In this case, the power per unit volume going into the Alfv\'en mode from fluid turbulence is
\begin{equation}
P_A=\int_{\bar{k}}^{\bar{k}_{\rm max}}I_A(\bar{k})d\bar{k}
\approx\frac{I_0 k_T}{(s_t-1)}\left(\frac{\bar{k}}{k_T}\right)^{(1-s_t)}\,,
\label{eq:PA1}
\end{equation}
where $\bar{k}\ll\bar{k}_{\rm max}$ and $s_t>1$.

On the other hand, according to the Lighthill theory, $P_A$ is given by (cf. Eq. (\ref{eq:PA}))
\begin{equation}
P_A=\eta_A\left[\frac{v(l)}{v_A}\right]\, \rho v^3(l)k\,,
\label{eq:PA2}
\end{equation}
where $\eta_A$ is an efficiency factor of order unity \citep{kato,hendr,eilek84}.
By comparing Eqs. (\ref{eq:PA1}) \& (\ref{eq:PA2}), and using Eqs. (\ref{6}) \& (\ref{7}), we obtain
\citep[see][]{eilek84,fuji},
\begin{equation}
s_t=\frac{3(m-1)}{(3-m)}\,,
\label{eq:s_t}
\end{equation}
\begin{equation}
I_0=\eta_A(s_t-1)\rho v_A^3\left(\frac{E_t}{\rho v_A^2R}\right)^{3/(3-m)}\,.
\label{eq:I_0}
\end{equation}

The power radiated in the form of Alfv\'en waves $P_A$ is dominated by small $\bar{k}$ and the smallest is $\bar{k}_T$.
With Eq. (\ref{vf}) and $\bar{k}_T=[v(l_T)/v_A]k_T$,
the total power in the form of Alfv\'en waves is approximately,
\begin{equation}
P_A=\eta_A\left(\frac{E_t}{\rho v_A^2R}\right)^2\rho v_A^3k_T\,.
\end{equation}
For Kolmogorov turbulence, the spectral index is $m=5/3$ (see Eq. \ref{eq:turbulence}),
and the spectral index of Alfv\'en wave flux is $s_t=3/2$ (Eq. \ref{ia}).

\section{Spectrum of MHD Turbulence in the Fermi Bubble Envelope}
\label{sec:MHD_turbulence}

The evolution of the spectrum of the Alfv\'en waves, $W_k(t)$ is described by the equation of nonlinear diffusion presented
in, e.g., \citet{brun04} and \citet{brun05},
\begin{eqnarray}
& {\displaystyle \frac{\partial W_k(\bar{k},t)}{\partial t}} & {\displaystyle =
\frac{\partial}{\partial\bar{k}}\left[D_{kk}\frac{\partial W_k(\bar{k},t)}{\partial\bar{k}}\right]} \nonumber\\
& &\quad {\displaystyle -\Gamma(\bar{k})W_k(\bar{k},t)+I_A(\bar{k},t)}\,.
\label{eq:spectrum_eq}
\end{eqnarray}

The first term on the right hand side of Eq. (\ref{eq:spectrum_eq}) describes the nonlinear MHD wave-wave cascade.
The diffusion coefficient for the Kolmogorov and the Iroshnikov-Kraichnan spectra
\citep[e.g.,][]{mill95}
\begin{eqnarray}
D_{kk} & = &
v_A \left\{
\begin{array}{l}
{\displaystyle \bar{k}^{7/2}\left[\frac{W_k(\bar{k},t)}{2W_B}\right]^{1/2}}\,,\ \mbox{Kolmogorov}\\
{\displaystyle \bar{k}^{4}\left[\frac{W_k(\bar{k},t)}{2W_B}\right]}\,,\ \mbox{Iroshnikov-Kraichnan}
\end{array}     \right.
\end{eqnarray}
where $W_B=B_0^2/8\pi$.

The second term on the right hand side of Eq. (\ref{eq:spectrum_eq}) describes the damping of MHD-waves by collisions of
relativistic and thermal particles in the interstellar or intercluster medium \citep[see][]{eilek79}
\begin{eqnarray}
& \Gamma_k &\simeq \frac{4\pi^3e^2v_A^2}{\bar{k}c^2}
\int_{p_{\rm min}}^{p_{\rm max}}p^2(1-\mu_\alpha)\frac{\partial F(p,t)}{\partial p}dp \nonumber\\
&&=\frac{\pi^2e^2v_A^2}{\bar{k}c^2}\int_{p_{\rm min}}^{p_{\rm max}}(1-\mu_\alpha) \nonumber\\
&&\quad\quad\quad\quad\quad\quad \times\left[\frac{\partial N(p,t)}{\partial p}-\frac{2N(p,t)}{p}\right]dp\,,
\label{eq:damping}
\end{eqnarray}
where $F(p,t)$ is the particle distribution function and $N(p,t)=4\pi p^2 F(p,t)$, and
\begin{equation}
\mu_\alpha=\frac{v_A}{c}\pm \frac{m\Omega}{p\bar{k}}\,.
\label{eq:mu_alpha}
\end{equation}
Here the upper and lower signs are for negative and positive charged particles, respectively.

The third term on the right hand side of Eq. (\ref{eq:spectrum_eq}), $I_A$,
describes the injection of Alfv\'en waves by the fluid turbulence through the Lighthill mechanism.

The time scale of the damping with the thermal pool is considerably shorter than the cascade time scale for
$\bar{k}/\bar{k}_{\rm max}\gg 0.1$.
Thus, a break or a cutoff in the wave spectrum is expected at large wavenumbers.

In the following, we present an alternative formulation of the steady-state equation for $W_k(\bar{k})$.
For simplicity it is described in a compact form \citep[see][and references therein]{norman96,ptus06}
\begin{equation}
\frac{\partial}{\partial\bar{k}}\left(\frac{\bar{k}W_k}{T_{\rm NL}}\right)=2\Gamma_{\rm CR}W_k + I_A(\bar{k})\,.
\label{eq:Wave_Eq}
\end{equation}
Here the source of MHD-fluctuations, $I_A(\bar{k})$, is given by Eq. (\ref{ia}).

The rate of damping MHD-waves by cosmic rays, $\Gamma_{\rm CR}(\bar{k})$, is \citep[see, e.g.,][]{ber90}
\begin{equation}
\Gamma_{\rm CR}(\bar{k})=\frac{\pi Z^2e^2v_A^2}{2\bar{k}c^2}\int_{p_{\rm res}(\bar{k})}^\infty\frac{dp}{p}F(p)\,,
\label{grc}
\end{equation}
where $F(p)$ is the CR distribution, $p$ is the particle momentum, and $p_{\rm res}(\bar{k})=ZeB/c\bar{k}$.

In the form of the Iroshnikov-Kraichnan spectrum, the term of the left hand side of Eq. (\ref{eq:Wave_Eq}) is
\begin{equation}
\frac{\partial}{\partial\bar{k}}\left(\frac{\bar{k}W_k}{T_{\rm NL}}\right)
=\frac{d}{d\bar{k}}\left[\frac{C\left(\bar{k}^{3}W^2(\bar{k})\right)}{\rho v_A}\right]\,,
\end{equation}
where the interaction is
\begin{equation}
T_{\rm NL}^{-1}({\bar k})=C_{\rm NL}\frac{\bar{k}^2W_k({\bar k})}{m_{\rm i}n_{\rm i} v_{\rm A}}\,,
\end{equation}
and the constant $C_{\rm NL}\sim 1$.

If the magnetic field fluctuations are injected by an external source at the scale $L=1/\bar{k}_L$,
Eq. (\ref{eq:Wave_Eq}) can be simplified to
\begin{equation}
\frac{\partial }{\partial\bar{k}}\left(\frac{\bar{k}W_k}{T_{\rm NL}}\right)=2\Gamma_{\rm CR}W_k
+ \Phi\,\delta(\bar{k}-\bar{k}_L)\,.
\label{eq:Wk}
\end{equation}
This result has been applied to the spectrum of MHD-turbulence in the FB envelope for $\bar{k}>\bar{k}_0$ \citep[][]{cheng14}.
The solution of Eq. (\ref{eq:Wk}) is given by
\begin{eqnarray}
&& W_k(\bar{k})=\left(\frac{\bar{k}_0}{\bar{k}}\right)^{3/2}W_k(\bar{k}_0) \nonumber\\
&&\quad -\,\frac{Z^2e^2B^2v_A}{8Cc^2 \bar{k}^{3/2}}\int_{\bar{k}_0}^{\bar{k}}{\bar k}_L^{-5/2}\,d\bar{k}
\int_{p_{\rm res}(\bar{k}_L)}^\infty\frac{F(p)dp}{p}\,,
\label{eq:Wk1}
\end{eqnarray}
where $W_k(\bar{k}_L)=\bar{k}_L^{-3/2}\sqrt{\rho v_A\Phi/C\,}$, where $\Phi$ describes the source injection at $\bar{k}_L$.

The coefficient of momentum diffusion of CRs is \citep[see][]{ber90}
\begin{equation}
D_p(p)=p^2\kappa(p)\,,
\label{eq:D_p}
\end{equation}
where
\begin{equation}
\kappa(p)=\frac{3\bar{k}_{\rm res}^2W_k(\bar{k}_{\rm res})}{\rho v}\,.
\label{kappa1}
\end{equation}
Here $\bar{k}_{\rm res}=1/r_L=ZeB/pc$, $r_L$ is the particle Larmor radius, $B$ is the magnetic field strength.

\begin{figure}[ht]
\centering
\plotone{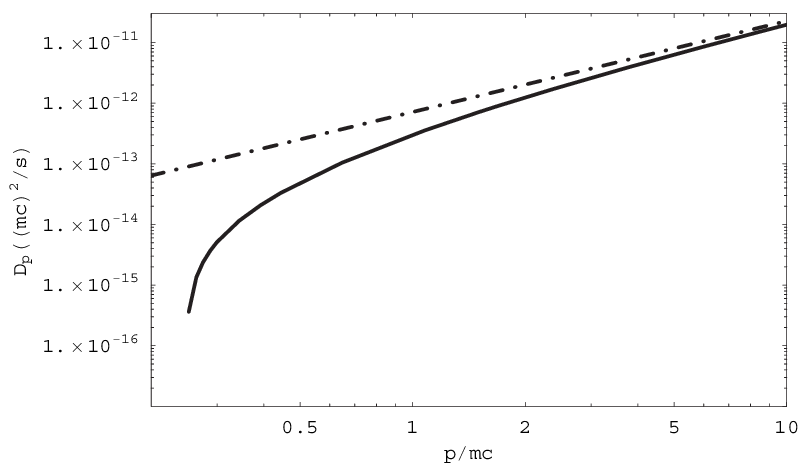}
\caption{
The solid line shows the momentum diffusion coefficient derived for the bubble parameters
when the CR absorption is taken into account.
The dash-dotted line is the results ignoring the CR absorption.
Figure reproduced from \citet{cheng14} with permission.
}
\label{Dp}
\end{figure}

The momentum diffusion coefficient $D_p$ for the Bubble parameters
is shown in Fig. \ref{Dp} (solid line). For comparison the dash-dotted line
is the diffusion coefficient for the Kraichnan spectrum of
turbulence without CR absorption.

As shown in Fig. \ref{Dp}, the wave damping by cosmic rays can terminate the cascade for relatively small CR momenta $p$.
\citet{brun04} showed that the time of damping  was considerably shorter than the cascade time for large wavenumbers.
They concluded also that the damping rate on protons largely dominates that on electrons.
These protons can exhibit a resonance with the relativistic electrons which may be important for their acceleration.

From Eqs. (\ref{eq:Wk1}), the distribution function of CR electrons, $F(p)$, can be estimated from the observed gamma-ray
\citep{su10,acker14} and microwave \citep{planck13} emissions for the expected parameter of FB envelope of hydrodynamic turbulence.

If the effect of CR damping is insignificant and $2\Gamma_{\rm CR}W_k$ can be ignored in Eq. (\ref{eq:Wave_Eq}),
then Eq. (\ref{eq:Wave_Eq}) is simply the balance of wave cascading and MHD-excitation by the hydrodynamic turbulence,
\begin{equation}
\frac{\partial }{\partial \bar{k}}\left(\frac{\bar{k}W_k}{T_{\rm NL}}\right)\approx I_A(\bar{k})\,.
\end{equation}
With Eq. (\ref{ia}), the spectrum of MHD-fluctuations $W_k$ is
\begin{equation}
W_k^2(\bar{k})\approx \frac{\rho v_A}{C_{\rm NL}\bar{k}^3}
\left[{\cal P}_0-\frac{I_0 k_0}{(s_t-1)}\left(\frac{k_0}{\bar{k}}\right)^{(s_t-1)}\right]\,,
\label{wbar}
\end{equation}
where
\begin{equation}
{\cal P}_0=\frac{I_0 k_0}{(s_t-1)}+\frac{C_{\rm NL}}{\rho v_A} k_0^3 W_k^2(k_0)\,.
\end{equation}
A cutoff in the spectrum of the waves $W_k$ can be estimated from the balance between damping and the cascade at large wavenumbers.

In contrast to the classical GALPROP code with accepted arbitrarily parameters of the kinetic diffusion
\citep[see, e.g.,][]{mos98,port08,vlad11},
we derive the coefficients of the kinetic equation for CRs in the FB envelope from Eq. (\ref{eq:Wk}) for the MHD-turbulence,
\begin{eqnarray}
&&\frac{\partial F(p)}{\partial t}+\frac{F(p)}{\tau_{\rm esc}}-Q(p,z) \nonumber\\
&=&\frac{1}{p^2}\frac{\partial}{\partial p}p^2\left[D_{p}(p)\frac{\partial F(p)}{\partial p}
-\left(\frac{dp}{dt}\right)F(p)\right] \nonumber\\
&&\quad\quad\quad\quad +\frac{\partial}{\partial z}D_{zz}(p)\frac{\partial F(p)}{\partial z}\,,
\label{aaa}
\end{eqnarray}
where $(dp/dt)$ ($>0$) is the rate of continuous energy losses,
$\tau_{\rm esc}$ is catastrophic CR losses or the characteristic time of CR escape from the envelope,
$Q(p,z)$ is the internal sources of CRs,
$D_{zz}$ is the coefficient of spatial diffusion
\begin{equation}
D_{zz}(p)=\frac{vB^2}{6\pi^2k^2W_k({\bar k})}=\frac{2\rho v_A^2 v}{3\pi\bar{k}^2W(\bar{k})}\,,
\label{eq:Dzz}
\end{equation}
and $D_{p}$ is the coefficient of momentum diffusion and is described by Eqs. (\ref{eq:D_p}) and (\ref{kappa1}).
Here $\bar{k}=1/r_L=ZeB/pc$ (cf. Eqs. (\ref{eq:D_p}) \& (\ref{kappa1})).

Our goal is to derive the spectrum of CRs from a combination of kinetic MHD/CR equations
and to estimate the proper and correct coefficients of Eq. (\ref{aaa}).
However, there is still a gap between the correct coefficients of the kinetic equations
and some rough estimations of the spatial and momentum diffusions from the observed gamma-ray and microwave emissions from the FBs.
At present we are unable to derive reliable numerical values of these coefficients,
and try to estimate roughly these parameters from the data ignoring the equation for the origin of MHD turbulence needed for
CR scattering and propagation.
These parameters of the spatial and momentum diffusion coefficients were roughly or arbitrarily estimated,
e.g., by weak random waves of a hydromagnetic turbulence \citep[see][]{sarkar},
by a supersonic turbulence \citep[see][]{byk93},
or by simple estimations of electron acceleration from shocks of the FBs \citep[see][]{cheng11}, etc.

In the following, we describe how to roughly estimate the parameters of the spatial and momentum diffusion coefficients
from the observed gamma-ray and microwave emissions from the FBs.

\section{Leptonic and Hadronic Origins of the Radiation from the Fermi Bubbles}
\label{sec:lepton_hadron}

The origin of CRs in the envelope of the giant bubbles is still an open question.
The structure of the bubbles is complicated.
It is seen in thermal X-ray as an outer envelope with the parameters:
\begin{eqnarray}
&&\mbox{Power of hydrodynamic turbulence}\ \sim 10^{39}\ \mbox{erg s$^{-1}$} \nonumber\\
&&\mbox{Thickness of the envelope}\ \sim 100\ \mbox{pc} \nonumber\\
&&\mbox{Scale of eROSITA bubbles}\ \sim 14\ \mbox{kpc} \nonumber\\
&&\mbox{Magnetic field}\ \sim 8\times 10^{-6}\ \mbox{G}\nonumber\\
&&\mbox{Gas density}\ \sim 4\times 10^{-3}\ \mbox{cm$^{-3}$}\nonumber\\
&&\mbox{Alfv\'en velocity}\ \sim 3\times 10^7\ \mbox{cm s$^{-1}$}\nonumber
\end{eqnarray}

\begin{figure*}[ht]
\centering
\plotone{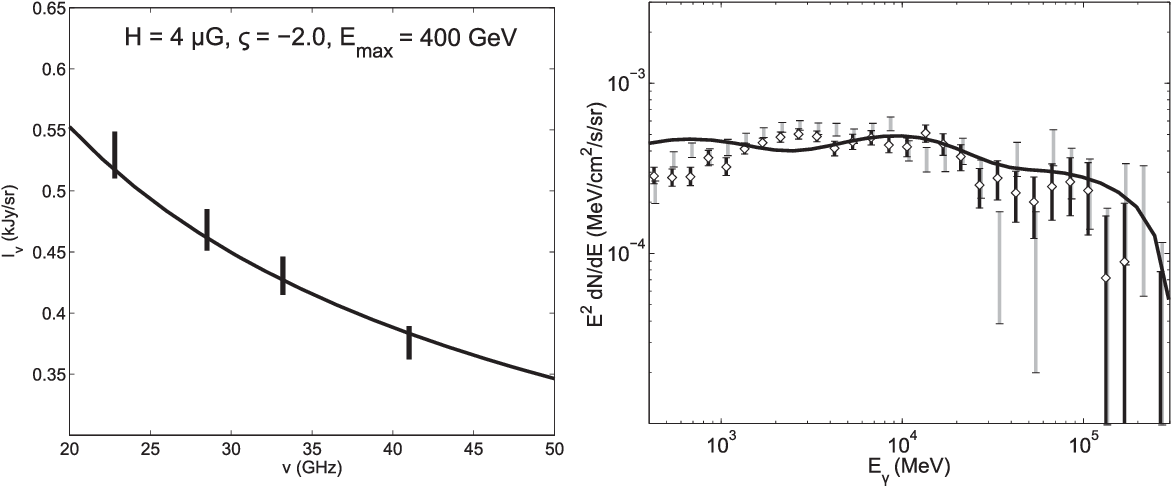}
\caption{
Spectrum of radio (left) and gamma-ray (right) emission from the FBs \citep[see][]{cheng14}.
The microwave data was taken from \citet{planck13}, and gamma-ray from \citet{acker12,acker14}.
Figure adapted from \citet{cheng14} with permission.
}
\label{Fermi-2}
\end{figure*}

An inner envelope of size about $\sim 8$ kpc is seen in the nonthermal gamma-ray and microwave emissions
(see Fig. \ref{Fermi-2}).
The microwave emission is evidently produced by the synchrotron losses of relativistic electrons,
while the origin of gamma-rays is not clear.

The total gamma-ray luminosity of the bubbles between 100 MeV and 500 GeV is
$F_\gamma\simeq 4.4  \times 10^{37}$ erg s$^{-1}$ \citep[see][]{acker14}.
The spectrum can be described by a power-law $dF_\gamma/dE_\gamma\propto E_\gamma^{-1.87}$
with a cut-off $E_{\rm cut}\simeq 113$ GeV.

\subsection{Origin of gamma-ray emission from the Fermi Bubbles}

The spectrum of gamma-rays can be fitted by either leptonic or hadronic model.
\begin{itemize}
\item Leptonic model: The rate of gamma-rays production by relativistic electrons
interacting with the low-energy interstellar photons is given by \citep[see][]{acker14}
\begin{eqnarray}
&&\varepsilon_{\rm IC}(E_\gamma)=c\sum_in_i(\epsilon_{\rm ph}) \nonumber\\
&&\quad\quad \times\int\frac{d\sigma_{\rm IC}(E_\gamma,E_e,\epsilon_{\rm ph})}{dE_\gamma} N_e(E_e)dE_e\,,
\end{eqnarray}
where $\sigma_{\rm IC}$ is the inverse Compton (IC) cross-section \citep[see, e.g.,][]{mos06}.
The parameters of the CR electron spectrum was derived from the observed gamma-ray emission from the envelope:
$N_e\propto E_e^{-2.17}$, $E_{\rm cut}\sim 1.25$ TeV.
The required total energy in electrons above 1 GeV is ${\cal E}_e\sim 10^{52}$ erg.
\item Hadronic model: gamma-ray can be produced by proton-proton (p-p) collisions.
For the calculations of the emission from p-p collision, \citet{acker14} used the p-p cross-section from \citet{kamae}.
The rate of gamma-ray production by p-p collisions is
\begin{equation}
\!\!\varepsilon_{\rm pp}(E_\gamma)=c\int\frac{d\sigma_{\rm pp}(E_\gamma,E_p)}{dE_\gamma}n_HN_p(E_p)dE_p\,.
\end{equation}
The required spectrum of CR protons is expressed as $dN_p/dE_p\propto E_p^{-2.13}\exp(-E_p/E_{\rm cut})$, where $E_{\rm cut}\sim 14$ TeV.
The needed total energy in CR protons above 1 GeV is ${\cal E}_p\sim 3.5\times 10^{55}$ erg for $n_H=0.01$ cm$^{-3}$.
\end{itemize}

In principle, we can interpret the gamma-ray emission (the right panel of Fig. \ref{Fermi-2}) by both leptonic (IC)
and hadronic (p-p) models for the correspondingly derived parameters of CRs.
The question is whether the observed microwave spectrum from the FBs is also compatible with the leptonic or hadron model.

\subsection{Microwave origin in cosmic ray electron model}
\label{sec:microwave_electron}

The origin of the microwave (see left panel of Fig. \ref{Fermi-2}) was analysed for leptonic and hadronic models \citep[see][]{acker14},
and the goal is to fit the gamma-ray and microwave observations within the same model.

The electron in the IC scenario should also produce the observed WMAP and Planck microwave spectrum and flux.
Their properties can be derived from the observed density of gamma-ray produced by relativistic electrons
that interact with the low-energy interstellar photons in the IC scenario for a magnetic field in the FBs
in the range between 5 $\mu$G and 20 $\mu$G \citep[see][]{acker14}.
The best-fit magnetic field is about 8.4 $\mu$G.
The synchrotron flux of the FBs from these electrons can be estimated as \citep[see][]{syr64,gs65},
\begin{eqnarray}
& \Phi_\nu &\simeq 4\pi\frac{\sqrt{3}e^3}{m_ec^2}\int_{r_{min}}^{r_0}B(r)r^2\,dr \nonumber\\
&&\quad\times\int\frac{\nu}{\nu_c} N_e(E)\,dE \int_{\nu/\nu_c}^{\infty}K_{5/3}(\eta)\,d\eta\,,
\label{Fnu}
\end{eqnarray}
where $K_\mu(\eta)$ is the McDonald function, and
\begin{equation}
\nu_c(r,E)=\frac{3eB(r)}{4\pi m_ec}\left(\frac{E}{m_ec^2}\right)^2\,.
\end{equation}

$\Phi_\nu(\nu(E_e))$ can be estimated from the density of relativistic electrons $N_e(E_e)$ in the FB envelope as,
\begin{equation}
N_e(E_e)\simeq \frac{\chi}{V_{0}}\frac{\Phi_\nu(\nu(E_e))}{E_e}\,,
\end{equation}
where the parameter $\chi$ is
\begin{equation}
\chi=\frac{3 m_e^3 c^5\nu_0}{e^4 B_0^2}\,.
\end{equation}

Although the main contribution to the IC signal comes from electrons at energies $>100$ GeV
while the main contribution to the Planck frequencies comes from electrons between $10\sim 30$ GeV, they match each other in the IC model.

\subsection{Microwave origin in cosmic ray proton model}
\label{sec:microwave_proton}

In the pure hadronic model \citep[see][]{acker14,cheng15a}, the FB synchrotron emission
is produced by secondary electrons from collisions of primary protons.

The approximated equations for the spectrum of secondary electrons produced by p-p and knock-on (KO) collisions ($pe$ and $ee$),
can be written as \citep[see, e.g.,][]{haya69,gin89}
\begin{eqnarray}
&&N_{\rm se}(E_e,r)= \nonumber\\
&&\quad\tau_e\int_{\frac{m_p}{m_e}E_e}^\infty N_p(E_p,r)n_H v_p d\sigma_{\rm pp}(E_p)\,, \\
&&N_{\rm ss}(E_e,r)=\nonumber\\
&&\quad\tau_e\int_{\frac{m_p}{m_e}E_e}^\infty N_p(E_p,r) v_p n_H \frac{d\sigma_{\rm KO}(E_p,E^\prime)}{dE^\prime}dE_p\,,
\end{eqnarray}
where $\tau_e$ is an integral over the rate of electron energy losses
\begin{equation}
\tau_e\sim\int_{E_e}^{E_{\rm max}}\frac{dE_e}{(dE_e/dt)_i}\,.
\end{equation}
Here $(dE_e/dt)_i$ can be determined by bremsstrahlung, ionization, synchrotron or particle escape.

The synchrotron emission of the secondary electron from the FB envelope can be calculated from Eq. (\ref{Fnu}).

The pure hadronic model is unable to reproduce both gamma-ray and radio fluxes from the FBs at the same time.
The problem is that the secondary electrons and positrons in the hadronic scenario
produce synchrotron radiation with a spectrum that is too soft compared to the microwave haze spectrum,
whereas the overall normalization of the synchrotron radiation from the secondary particles
is at least a factor of three to four smaller than the microwave level that a hadronic model requires
\citep[see][]{cheng15a}.

Thus, we conclude that a purely hadronic origin of the nonthermal emission (gamma and radio) from the FBs is problematic.

\section{Number of Relativistic Electrons in the Fermi Bubble Envelope}
\label{sec:relativistic_electron}

The origin of relativistic electrons in the FB envelope is an open question.
It was assumed that the FB envelopes might be bounded by a shock with the velocity $v_{\rm sh}\sim 10^8$ cm s$^{-1}$
\citep[see, e.g.,][]{cheng11,dorf12,per22}.
It was proposed that these CRs are accelerated at the shock by the standard mechanism of shock acceleration
with the CR spectrum $Q(E_e)\propto E_e^{-2}$ \citep[see, e.g.][]{axford,krym,bell,bland}.
However, the eROSITA \citep[see][]{predehl} found that X-ray giant bubbles propagate with the velocity of the shock
is about $340$ km s$^{-1}$  and its Mach number is only $\approx 1.5$, which does not correspond to an effective CR shock acceleration.
Thus, CRs are not accelerated by a shock near the outer shell of eROSITA.
Therefore, CRs should be produced by in-situ stochastic acceleration by MHD-turbulence $W_k$ nearby the inner Bubble surface
(see Section \ref{sec:MHD_turbulence} the function $W_k$).

In the following, we focus on the in-situ stochastic (Fermi) acceleration of CRs by a hydromagnetic or supersonic turbulence.

\subsection{Electrons accelerated from background plasma in the Fermi Bubbles}
\label{sec:electron_acc}

In the model of \citet{cheng14}, CR electrons can be directly accelerated from a background plasma.
They suggested that the acceleration from the background plasma is able to explain the origin of the nonthermal particles
responsible for producing the observed fluxes of radio and gamma-ray emissions from the bubbles.

The kinetic equation for the distribution function of electrons, $F(p,t)$, in the case of in-situ acceleration is described as
\begin{eqnarray}
&&\frac{\partial F(p,t)}{\partial t}+\frac{F(p,t)}{\tau_{\rm esc}} \nonumber\\
&=&\frac{1}{p^2}\frac{\partial}{\partial p}p^2\left[-\left(\frac{dp}{dt}\right)_{\rm C} F(p,t) \right. \nonumber\\
&&\quad\quad\quad\quad \left.+\left\{D_{\rm C}(p)+D_{\rm F}(p)\right\}\frac{\partial F(p,t)}{\partial p}\right]\,.
\label{eq_nr}
\end{eqnarray}
The distribution function includes the thermal and nonthermal components of the particle distribution.
Coefficient $(dp/dt)_{\rm C}$ describes particle ionization/Coulomb energy losses.
$D_{\rm C}(p)$ describes diffusion in the momentum space due to Coulomb collisions \citep[for detail see][]{LP}.
The stochastic (Fermi) acceleration is described as diffusion in the momentum space with the diffusion coefficient $D_{\rm F}(p)$.
$\tau_{\rm esc}$ is the lifetime of particles in the region of acceleration, e.g., due to escape from the region.

In an ionized plasma, the equilibrium (Maxwellian) spectrum of background charged particles is formed by Coulomb collisions.
For the case of CR acceleration, there is a  boundary, $E=E_{\rm inj}$ between the equilibrium Maxwellian distribution
and a power-law nonthermal spectrum of accelerated particles,
\begin{equation}
\frac{dE}{dt} =\alpha_0 E-\nu_0 E\left(\frac{kT}{E}\right)^{3/2}\,.
\end{equation}
The  energy of injection is
\begin{equation}
E_{\rm inj}\sim  kT \left(\frac{\nu_0}{\alpha_0}\right)^{2/3}\,.
\end{equation}
Here the parameters $\alpha_0$ and $\nu_0$ are the acceleration and the ionization loss (by Coulomb collisions), respectively.

Making use of Eq.~(\ref{eq_nr}), \citet{gur60} studied the process of the formal connection between
equilibrium Maxwellian distribution of background particles
and a power-law non-equilibrium spectrum of accelerated particles.

For slow time variations, the run-away flux of particles from the region of thermal Maxwellian distribution
into the acceleration range is \citep[see, e.g.,][]{dog00}
\begin{equation}
S_\zeta=S_0(t)\sqrt{\frac{2}{\pi}}\int_0^\zeta \zeta^2\exp\left(-\frac{\zeta^2}{2}\right)d\zeta\,,
\label{eq:S_zeta}
\end{equation}
where $\zeta=p/p_0$ is the normalised momentum, and
\begin{equation}
S_0(t)=\sqrt{\frac{2}{\pi}}n(t)\exp\left(-\int_0^\infty \frac{\zeta\,d\zeta}{(1+\alpha_0 \zeta^5/\nu_0)}\right)\,,
\label{eq:S_0}
\end{equation}
and $n(t)$ is the slow variation of the gas density.

In a non-equilibrium case this process forms an escape flux of runaway particles for energy $E>E_{\rm inj}$.
On the other hand, this process also forms an excess density at energies below $E\lesssim E_{\rm inj}$,
which distorts the thermal Maxwellian distribution.
This can be interpreted as a ``second'' effective temperature higher than the gas equilibrium temperature.


This model of \citet{gur60} has been applied to the processes of particle acceleration from background plasma in galaxy clusters
\citep[see][]{dog00,liang02,dog07}, in the GC \citep[see][]{cheng14}, and in the Galactic disk \citep{dog02},
in which excesses above thermal particles in X-ray range is expected.

\begin{figure}[ht]
\centering
\plotone{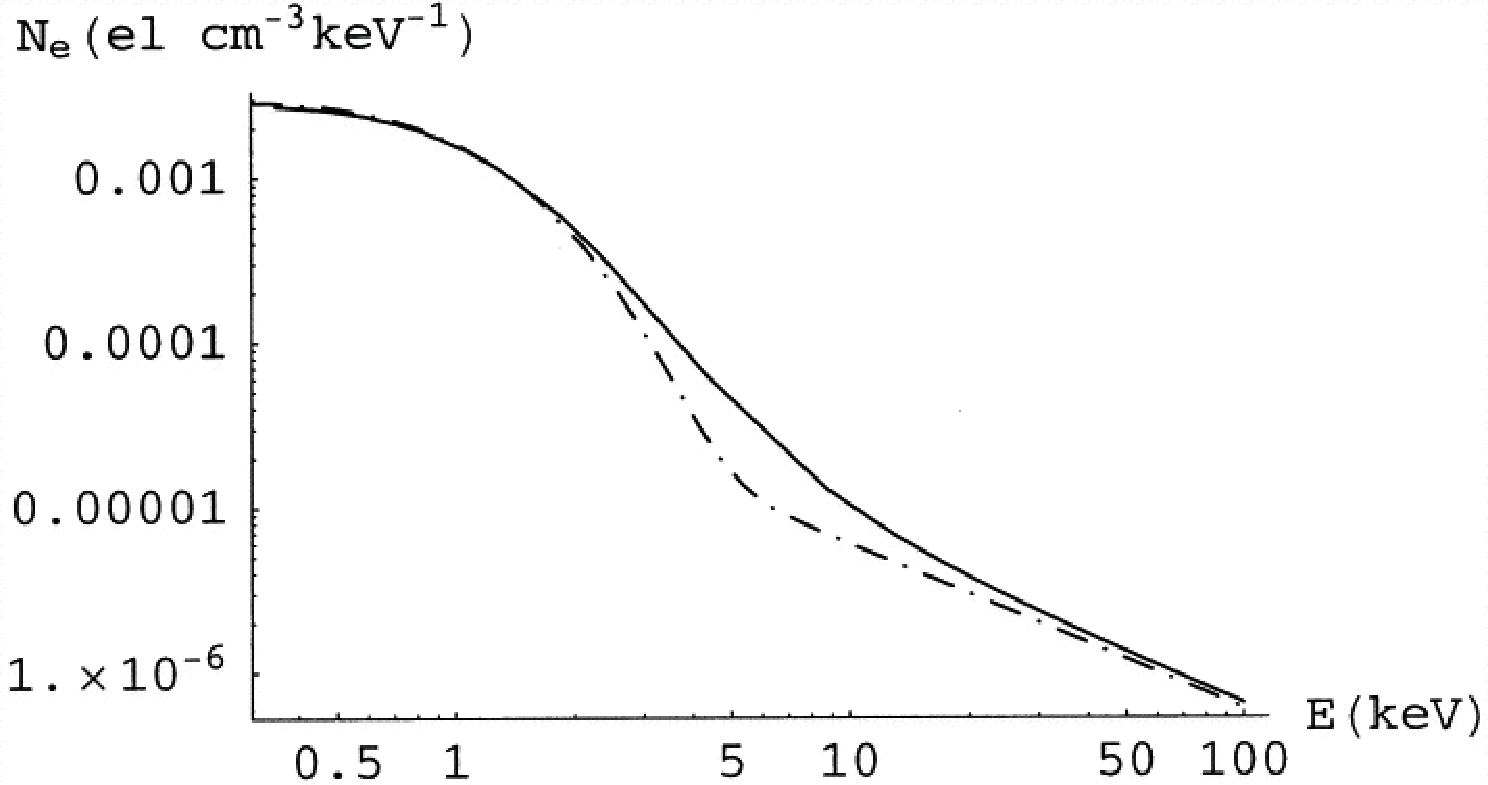}
\caption{
X-ray emission from the Galactic plane whose excess emission is above the equilibrium Maxwellian spectrum.
Dash-dotted line is a simple combination of thermal plus nonthermal spectrum.
Solid line is the spectrum with the effect of runaway flux.
Figure reproduced from \citet{dog02} with permission.
}
\label{St}
\end{figure}

For example, the spectrum of X-ray emission from the Galactic plane can be described as a multi-temperature emission.
\citet{dog02} interpreted this X-ray emission as the flux of runaway particles from background gas as
a common effect of Coulomb collision (ionization losses) and stochastic acceleration (see Fig. \ref{St}).
When compared with the spectrum of a simple combination of thermal and nonthermal gas, the spectrum with runaway flux is
larger, in particular, in the transition range between the thermal and nonthermal part.

The model of particle excess from background gas proposed for CRs by \citet{dog00} was challenged by \citet{petr01}, \citet{wolfe06} and \citet{east08}.
The problem was that the stochastic acceleration from a background plasma would (over) heat the plasma by the accelerated particles,
because their energy would be quickly dumped into the thermal plasma by ionization losses.
The energy, gained by the particles, is distributed to the whole plasma on a timescale much shorter than that of the acceleration process itself.
As a result of the relatively inefficiency of bremsstrahlung for cooling the accelerated electrons,
this tail is quickly dumped into the thermal body of the background plasma (plasma overheating without a prominent tail of accelerated particles).
This effect prevents completely the formation of nonthermal spectra from background plasma.

\begin{figure}[ht]
\centering
\plotone{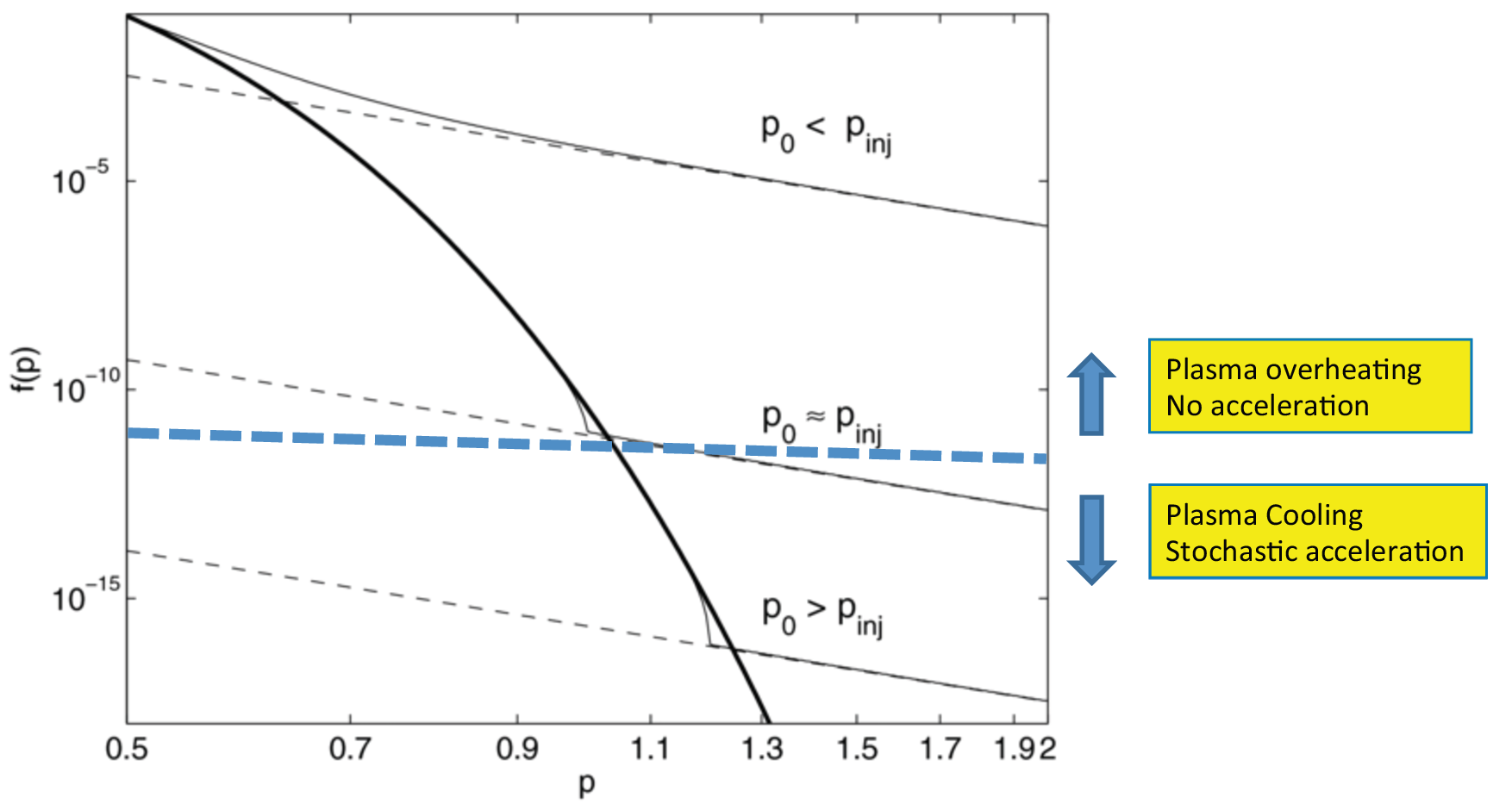}
\caption{
The spectrum of electrons accelerated from background plasma \citep[see][]{chern12}.
The solid line is the density of electrons, $f(p)$.
The thick solid line is the pure thermal Maxwellian distribution.
The dashed line is the power-law approximation of the nonthermal tail.
For $p_0>p_{\rm inj}$, overheating is insignificant.
Figure adapted from \citet{chern12} with permission.
}
\label{inj}
\end{figure}

However, \citet{chern12} showed that the effect of overheating depends on the parameters of acceleration.
It is insignificant if the stochastic acceleration is effective.
This model depends on a value of $p_{\rm inj}$ and a free parameter of stochastic acceleration $p_0$ in the form of
\begin{equation}
D_{\rm F}(p)=D_0\,p^\varsigma\theta(p-p_0)\,,
\end{equation}
where $D_0$ and $\varsigma$ are constants. In general, $p_{\rm inj}$ is determined by $D_{\rm F}(p_{\rm inj})=[p(dp/dt)_{\rm C}]_{p_{\rm inj}}$.
In this model, the injection momentum is given by
\begin{equation}
D_0\,p_{\rm inj}^{(\varsigma-1)}=\left[\left(\frac{dp}{dt}\right)_{\rm C}\right]_{p_{\rm inj}}\,.
\end{equation}
For a high value of the acceleration momentum $p_0$ the runaway flux of thermal particles cools the plasma down from the very beginning.
In spite of energy supply by external sources the plasma temperature drops down (analogue to Maxwell demon).
Acceleration generates a prominent tail of accelerated particles but``excess'' is not produced at the range around $p_{\rm inj}$
that was expected in \citet{gur60}, see Fig. \ref{inj}.
For $p_0>p_{\rm inj}$ plasma overheating is insignificant and stochastic acceleration works well.
For $p_0<p_{\rm inj}$ plasma overheating is significant and stochastic acceleration is inhibited.

In any case, numerical calculations showed that the permitted parameters of the FB model are strongly restricted,
and the model is unable to explain the observed fluxes of radio and gamma-ray emissions from the bubbles \citep[see][]{cheng14}.

\subsection{Cosmic ray electrons re-accelerated in the Fermi Bubbles}
\label{sec:electron_reacc}

CR electrons can be generated by sources in the Galactic disk (e.g., supernova remnant shocks, SNR shocks).
Due to synchrotron, inverse Compton and adiabatic losses in the halo,
\citet{cheng15b} deemed that these CR electrons with energies above several GeV are unable to reach the height of the FB envelope
(which is about $8\sim 10$ kpc).
With appropriate parameters in the FB envelope, these electrons can be re-accelerated in-situ up to energies about $10^{12}$ eV,
which is needed to reproduce the observed radio and gamma-ray emissions from the FBs and supply the required power.

The steady state kinetic equation for the relativistic electrons in FBs can be described in the form \citep[see][]{cheng15b}
\begin{eqnarray}
&\!\!\!\!\!\!\!Q(p)\delta(z)&=-\nabla\cdot\left[D_s\nabla F-{\bf v}F\right] \nonumber\\
&& +\frac{1}{p^2}\frac{\partial}{\partial p}p^2
\left[\left(\frac{dp}{dt}-\frac{\nabla\cdot{\bf v}}{3}p\right)F-D_p\frac{\partial F}{\partial p}\right]\,,
\label{eq:steady}
\end{eqnarray}
where $F(p,r,z)$ is the particle distribution function,
$Q(p,r)\propto p^{-\gamma}$ is the source function of electrons in the Galactic disk ($z=0$),
$dp/dt=\mu E^2$ is the rate of synchrotron and inverse Compton energy losses,
$D_s(p,r,z)$ is the coefficient of spatial diffusion,
and $D_p(p,r,z)=\kappa p^2$ is the coefficient of momentum diffusion (coefficient of the Fermi re-acceleration),
and ${\bf v}(r,z) = v_z{\hat e}_z=3v^\prime z {\hat e}_z$ ($v^\prime=dv_z/dz$) is the wind velocity of advection in the halo in the $z$ direction.
The effect of the wind advection leads to adiabatic losses of CRs, $dp/dt=-p\nabla\cdot{\bf v}/3$.
Consequently, the spectrum of electrons in the FBs in the acceleration region is harder than the case without advection.

\citet{cheng15b} showed that the gamma-ray and radio emissions of the re-accelerated electrons reproduced nicely the Fermi-LAT
and Planck data points for the parameters:
the spatial diffusion coefficient $D_s=10^{29}$ cm$^2$ s$^{-1}$,
the energy loss rate $\nu=2\times 10^{-16}$ s$^{-1}$ GeV$^{-1}$,
the velocity gradient of the advection in the halo $v^\prime=10^{-15}$ s$^{-1}$,
the magnetic field strength is $B=3$ $\mu$G,
the thickness of the re-acceleration region (say, the FB envelope) is about $\Delta r_{\rm FB}=60$ pc,
and the parameter of re-acceleration in the FBs is $\kappa=2\times 10^{-14}$ s$^{-1}$.

The total power $\dot{\cal E}$ supplied by sources of Fermi re-acceleration in the FBs to produce high-energy electrons is needed
\begin{equation}
\dot{{\cal E}}=-\int_0^\infty 4\pi E\frac{\partial}{\partial p}\left(p^2 D_p\frac{\partial F}{\partial p}\right)dp\,,
\label{eq:power}
\end{equation}
where $p$ and $E$ are the particle momentum and particle kinetic energy, respectively.

\begin{figure}[ht]
\centering
\plotone{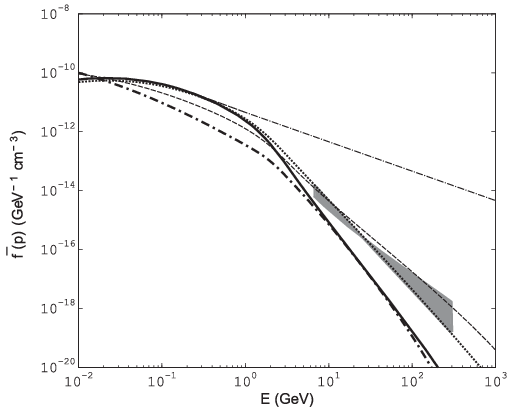}
\caption{
The spectrum of SNR electrons from the Galactic disk that have been re-accelerated in the FBs.
The five spectra in the figure correspond to different cases of the model:
(1) thick solid line: without re-acceleration, escape and advection;
(2) thick dash-dotted line: without re-acceleration and escape but with advection;
(3) thin dash-dotted line: with re-acceleration but without escape from the region and advection;
(4) thin dotted line: with re-acceleration and escape from the region but without advection;
(5) thin dashed line: with re-acceleration and advection but without escape.
The density of electrons needed for the observed gamma-ray flux from the bubbles is shown by the gray region.
The electron spectrum of case (5) can reproduce the gamma-ray data from Fermi-LAT and the microwave data from Planck (Fig. \ref{Fermi-2}).
The parameters of case (5) can be found in the main text.
For parameters of other cases, the reader is referred to \citet{cheng15b}.
Figure reproduced from \citet{cheng15b} with permission.
}
\label{Fig_dif_conv}
\end{figure}

To define the spectrum of accelerated electrons we estimated the number of GeV electrons that can reach an altitude of several kpc
when the effect of advection $v(z)$ is essential in the Galactic halo \citep[see][]{brei91,bloemen93,brei02,blasi12}.
The spectrum of re-accelerated SNR electrons in the FBs is shown in Fig. \ref{Fig_dif_conv} \citep[cf.][]{cheng15b}.
In the figure, the thick solid line is the spectrum of CR electrons from their sources in the Galactic disk \citep[see, e.g.,][]{ber90}.
When re-acceleration (stochastic acceleration) and adiabatic losses are taken into account,
the spectrum of electrons (thin dashed line) becomes harder than that of spectrum emitted by sources (thick solid line),
but softer than the spectrum of pure re-acceleration (thin dash-dotted line).
The spectrum for a velocity gradient $v^\prime =10^{-15}$ s$^{-1}$ \citep[see, e.g.,][]{bloemen93,brei02}
is consistent with the observed one (dashed line in Fig. \ref{Fig_dif_conv}).

The model of re-acceleration within the envelope coincides nicely with the observed microwave and gamma-ray emissions
shown in the left and right panels of Fig. \ref{Fermi-2}, respectively.

In the phenomenological model, the power, $\dot{{\cal E}}$, is estimated numerically from the observed FB gamma-ray and microwave fluxes,
and it is about $\dot{{\cal E}}\sim 2\times 10^{38}$ erg s$^{-1}$.
The density of high-energy electrons needed for the observed gamma-ray flux is shown by the shaded gray region in Fig. \ref{Fig_dif_conv}.

With appropriate parameters these electrons can be re-accelerated up to an energy of $10^{12}$ eV,
which explains in this model the origin of the observed gamma-ray and radio emissions from the FBs.
However, although the model gamma-ray spectrum is consistent with the Fermi results,
the model radio spectrum in the pure diffusion model is steeper than that observed by WMAP and Planck.

If adiabatic losses due to plasma outflows from the Galactic central regions are taken into account,
we expect that the spectrum of electrons in the acceleration region will be harder than the one without advection.
Our calculations with divergent outflows show that the gamma-ray and radio emissions by the re-accelerated electrons
nicely reproduce the Fermi-LAT and the Planck data
(see Fig. \ref{Fermi-2}).

In essence, both gamma-ray and microwave observations can be explained by only one source of high energy electrons.
The basic idea is summarized as follows.
CR electrons from SNR in the Galactic disk are re-accelerated in the FB via supersonic turbulence (or multiple shocks).
The resulting spectrum is hard, and in the high energy region the density of electrons exceeds
the one to produce the observed gamma-ray emission. With adiabatic loss by the divergent flow,
the density reduces but the spectrum is still hard enough to produce the observed microwave emission.
Inevitably, some delicate balance or fine-tuning of parameters is needed.

\section{Cosmic Ray Protons Escaping from the Fermi Bubbles into the Galaxy}
\label{sec:proton_galaxy}

The fundamental question of the sources of CRs in the galaxy is still open.
We present a number of models which may interpret the origin of CRs escaping from the FBs into the Galactic disk.
The effect can be observed from the spectrum of CRs near Earth.

As described in \citet{ber90}, the classical model of CR origin in the Galaxy is
CRs are generated by SNRs in the Galactic disk with energies below $10^{15}$ eV.
They escape into the Galactic halo with an effective spatial diffusion coefficient about
$D_{\rm G} = D_*(E/4\,{\rm GeV})^{0.6}$ ($D_* = 6.2 \times 10^{28}$ cm$^2$ s$^{-1}$),
which is estimated from the observed chemical composition of CRs
\citep[for a modern nonlinear model of CR see][and references therein]{dog20}.
Plane shock acceleration produces a CR spectrum $F(E)\propto E^{-2}$ \citep[see, e.g.,][]{axford,krym,bell,bland}.
The actual CR spectrum observed outside the acceleration region is a result of the process of acceleration
in the region together with the process of particle leakage or escape from the region.
Furthermore, the maximum energy that can be attained by CR particles depends on the size and/or lifetime of the acceleration region.
Suppose $v_{\rm sh}$ is the shock velocity, and $D_{\rm sh}$ is the spatial diffusion coefficient in the shock vicinity.
The minimum value of $D_{\rm sh}$ follows the Bohm's limit, $D_{\rm sh}\approx u r_L/3$ (where $u\approx c$ is the speed of CR particles).
The maximum energy is constrained by the size of the acceleration region becomes
smaller than the diffusion length scale of the particles
$l_{D}\sim D_{\rm sh}/v_{\rm sh}$,
and/or the lifetime of the region is smaller than the acceleration time scale of the particles $\tau_{\rm acc}\sim D_{\rm sh}/v_{\rm sh}^2$.

\begin{figure}[ht]
\centering
\plotone{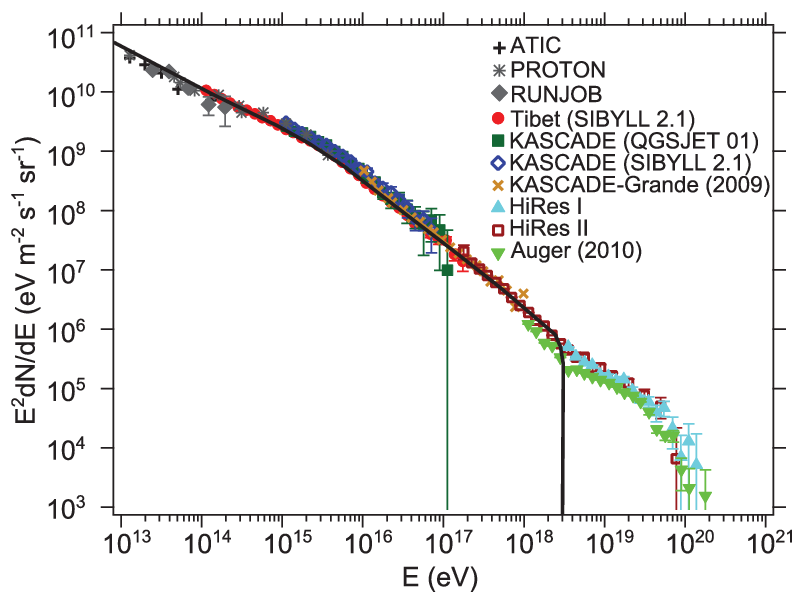}
\caption{
CR spectrum at the Earth as a combination of the contributions from the SNRs in the Galactic disk
and the stochastic acceleration in the FBs.
Figure reproduced from \citet{cheng12} with permission.
}
\label{comb}
\end{figure}

The observed CR spectrum (above $1$ GeV) at Earth can be described as a broken power law,
see the classic spectrum in \citet{Swordy2001} (or Fig. \ref{comb} which shows the part above $10^{13}$ eV).
The power law index is $-2.7$ between $10^9$ and $10^{15}$ eV, and $-3.1$ between $10^{15}$ and several $10^{18}$ eV.
Energy around $10^{15}$ eV is called the `knee', where the spectrum changes from a harder one to a softer one.
The spectrum beyond several $10^{18}$ eV becomes harder again and this region is called the `ankle'.
The apparent cutoff at somewhat less than $10^{20}$ eV is commonly attributed to the Greisen-Zatsepin-Kuzmin limit
due to the interaction of ultra high energy CRs with cosmic microwave background \citep[][]{Greisen1966,ZatsepinKuzmin1966}.
At energy smaller than $1$ GeV, the spectrum is heavily affected by solar modulation and activity of the Sun.
Although the spectrum is a broken power law, the `joints' (say the `knee' and the `ankle') are smooth. The origins
of different parts of the spectrum should be somehow related \citep[e.g.,][]{Axford1994}.

SNR shocks are believed to be the source of CRs \citep[e.g.,][]{ber90}. They can produce the spectral index $-2.7$ reasonably well.
With the magnetic field around the SNR comparable the general interstellar field,
\citet{Lagage1983} and \citet{Berezhko2000} estimated the maximum energy of protons from SNRs is about $10^{13}\sim 10^{14}$ eV.
However, with instabilities caused by cosmic ray streaming (e.g., non-resonant hybrid instability),
\citet{bell04} and \citet{byk09} found that the fluctuated magnetic field can be orders of magnitude larger at the SNR shock
(and so $D_{\rm sh}\ll D_{\rm G}$).
The confinement time of CRs is longer and they can be accelerated up to $10^{15}$ eV \citep[see also][]{Bell2013}.

The SNR shock framework is able to describe the spectrum in the energy range $E<10^{15}$ eV (see Fig. \ref{comb}).
However, for $E>10^{15}$ eV, we need some other ideas \citep[see, e.g.,][]{byk93}.
Larger and longer lifetime shocks are required, such as superbubbles
(FB is an example).
and strong galactic winds
\citep[e.g.,][]{Jokipii1985,Volk2004,cheng12,dorf12,per22,Zirakashvili2024}.
In the following, we analyze the spectrum of CRs accelerated in the FBs
and discuss whether the bubble's contribution may explain the `knee' steepening.
The idea is based on acceleration by multiple shocks.
In a multiple-shock system, two length scales are important:
(1) the average separation between two shocks $l_{\rm sh}\sim v_{\rm sh}\tau_{\rm c}$
(where $\tau_{\rm c}$ is the average time between the creation of two consecutive shocks),
and (2) the diffusion length scale at the shock
$l_D$ ($\sim D_{\rm sh}/v_{\rm sh}$).
We will focus on the regime $l_{\rm sh}\ll l_D$ (the supersonic turbulence regime),
for energies beyond the `knee' ($E>10^{15}$ eV).

\begin{figure}[ht]
\centering
\plotone{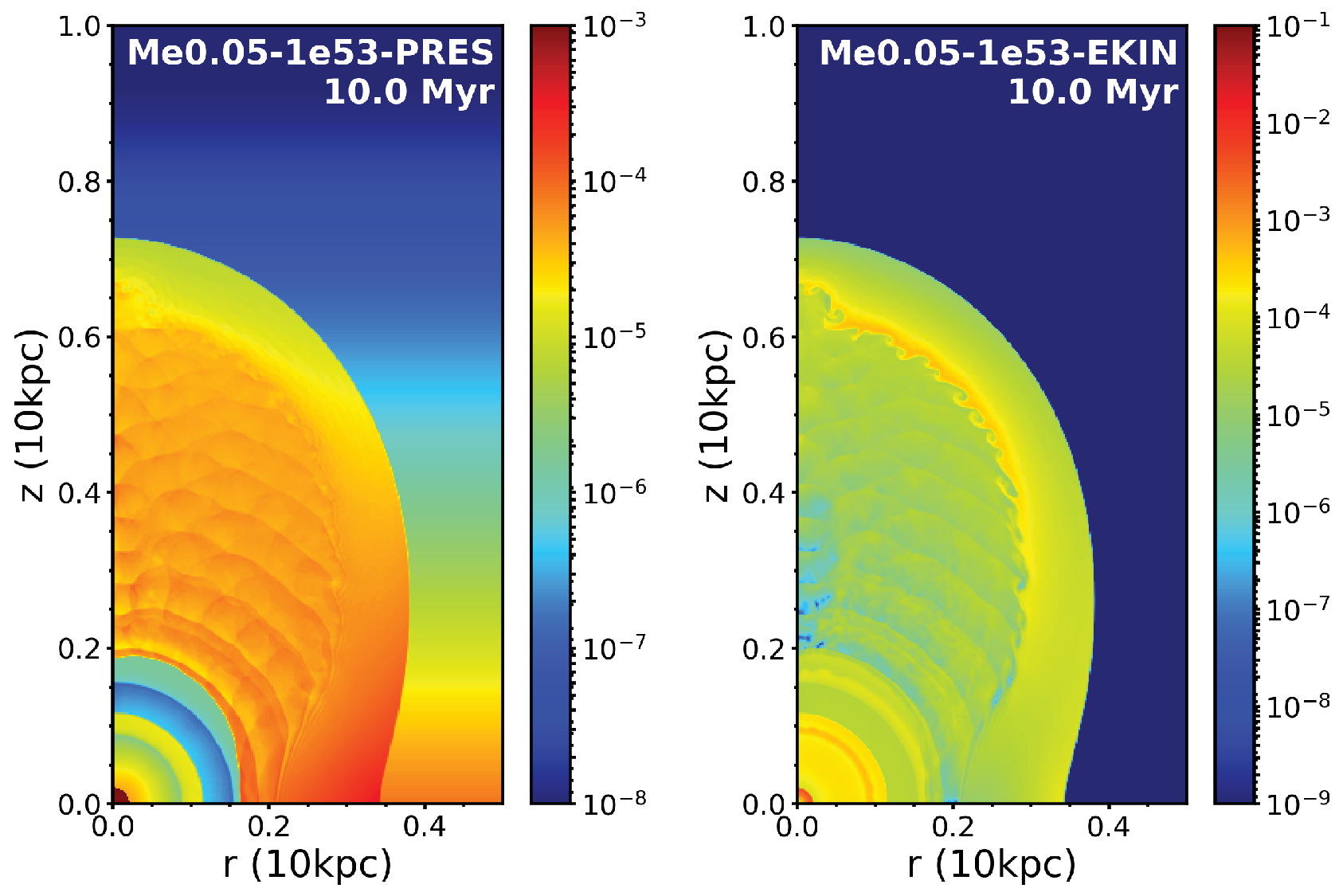}
\caption{
A possible multiple-shock structure in the FBs resulting from multiple TDEs at the GC.
The figure shows the pressure (left panel) and kinetic energy (right panel) distributions of a numerical simulation
of the FBs in an exponential halo.
In the panels, ``Me0.05-1e53'' corresponds to multiple TDEs with 0.05 Myr between successive TDE
and the energy release by TDE is $10^{53}$ erg.
The simulation ends at 10.0 Myr.
The unit of the color bar in both panels is $1.178\times 10^{-8}$ erg cm$^{-3}$.
}
\label{sturb}
\end{figure}

An illustration of a possible multiple-shock structure in the FBs is shown in Fig. \ref{sturb}.

\subsection{Escape of cosmic ray protons re-accelerated by supersonic turbulence inside the Fermi Bubbles}
\label{sec:proton_turbulence}

\citet{cheng12} suggested an alternative model of CRs in the FBs \citep[see also][]{cheng06,cheng07,dog09a,dog09b,dog09c}.
They assumed that up to several hundred TDEs might have occurred in the past 10 Myr.
This would have generated a series of shocks propagating through the central part of the Galactic halo,
which would produce relativistic CRs via multiple-shock acceleration.
The average separation between two shocks is then
\begin{eqnarray}
&l_{\rm sh}&=v_{\rm sh}\tau_{\rm cap} \nonumber\\
&&\simeq 30\left(\frac{\tau_{\rm cap}}{3\times 10^4\,{\rm yr}}\right)
\left(\frac{v_{\rm sh}}{10^8\,{\rm cm\, s}^{-1}}\right)\ {\rm pc}\,,
\end{eqnarray}
where $\tau_{\rm cap}$ is the average time between two stellar captures by the SMBH.

We applied the model of \citet{byk92} and \citet{byk93} for CRs acceleration by multiple shocks in the FBs.
Under the conditions of supersonic turbulence (multiple-shock structure) the regime of acceleration is characterized by $l_{\rm sh}\ll l_{\rm D}$.


The steady state kinetic equation in axisymmetric geometry is
\begin{eqnarray}
&&\frac{\partial}{\partial z}\left[D_s(p,r)\frac{\partial F}{\partial z}\right]
+\frac{1}{r}\frac{\partial}{\partial r}\left[D_s(p,r)\,r\frac{\partial F}{\partial r}\right] \nonumber\\
&&\quad +\frac{1}{p^2}\frac{\partial}{\partial p}\left[D_p(p,r)p^2\frac{\partial F}{\partial p}\right]=-Q(p,r,z)\,.
\label{2d}
\end{eqnarray}
The spatial diffusion coefficient inside and outside the bubble is,
\begin{equation}
D_s(p,r) = D_B\theta(r_{\rm B}-r)+D_{\rm G}\theta(r-r_{\rm B})\,,
\label{eq:diffcoeff}
\end{equation}
where $r_{\rm B}$ is the radius of the bubbles.
Inside the bubbles, as a result of interaction with supersonic turbulence $D_s=D_{\rm B}\approx u l_{\rm sh}/3\approx c l_{\rm sh}/3$.
Outside the bubbles, $D_s=D_{\rm G}$, is the average diffusion coefficient in the Galaxy, e.g., the one described in \citet{ber90}.
The momentum diffusion coefficient is nonzero inside the bubbles only,
\begin{equation}
D_p(p,r)=\kappa_{\rm B} p^2\theta(r_{\rm B}-r)\,,
\end{equation}
and $\kappa_{\rm B}\sim v_{\rm sh}^2/D_{\rm B}$.
The parameters in the FBs can be estimated numerically from the observed CR spectrum.


Using the method of separation variables as in \citet{bul72} and \citet{bul74} to solve Equation~(\ref{2d}) for the spectrum of CRs,
generated by SNRs with the standard model of CR propagation and escape in the Galactic halo \citep[see, e.g.,][]{syr64,ber90,Strong2007}.
This model describes the observed CR spectrum near Earth in the range below the `knee', $10^{15}$ eV (see Figure~\ref{comb}).

\citet{cheng12} interpreted the CR spectrum near Earth in the energy range from $10^{15}$ eV to a few $10^{18}$ eV
(from the `knee' to the `ankle') as a combined result of acceleration in the FB and escape.
The acceleration is provided by the supersonic turbulence in the FB \citep[see][]{byk93}.
Under the set of parameters in \citet{cheng12}, the ratio of the escape time to the acceleration time is about 1.9.

In order to derive the CR spectrum near Earth, \citet{cheng12} matched the solutions for the spectra inside the FB and outside (in the halo)
at $r=r_{\rm B}$ (see Equations (\ref{2d}) and (\ref{eq:diffcoeff})).
Basically, the model can reproduce the CR spectrum from $10^{13}$ eV to several $10^{18}$ eV.
However, we should point out that the numerical result described here \citep[][]{cheng12} has free parameters and some physics have been
ignored (e.g., adiabatic loss), and the result might not be very solid. It is imperative to have further investigations.

A brief summary of the idea is as follows.
CRs from SNR ($<10^{15}$ eV) are re-accelerated in the FB by multiple shocks or supersonic turbulence.
As the shocks are larger and live longer in FB, CRs can be accelerated to much higher energies (up to $10^{19}$ eV).
Acceleration by multiple shocks inside the bubbles gives a harder spectrum,
hence the contribution of the FB to CRs within the `knee' ($<10^{15}$ eV is subordinate to SNR).
The high energy CRs escaping from the bubbles constitute the sole source of CRs in the range between the `knee' to the `ankle'
($10^{15}$ to several $10^{18}$ eV) observed at Earth.
In the model, some fine-tuning of parameters is inevitable.

\section{Summary}
\label{sec:summary}

Here we present a brief summary of our perspective of the Fermi Bubbles at the Galactic Centre.

\begin{itemize}
\item
The key point of the bubbles is a huge energy release of $10^{55}\sim 10^{56}$ erg in the Galactic centre
whose origin is still unknown yet.
\item
We assume that the energy source of the bubbles could be a routine tidal disruption of stars near the central supermassive black hole.
Each disruption of a star releases a total energy about $10^{52}\sim 10^{53}$ erg.
For a typical rate of stellar capture (once $10^4$ years),
this can provide a luminosity $\gtrsim 10^{41}$ erg s$^{-1}$ from the Galactic Centre.
These processes of stellar tidal disruption events can be directly observed in some external galaxies.
\item
Hydrodynamic models can describe the envelope of bubble propagation in the Galactic halo where the gas distribution is nonuniform.
The distribution is commonly characterized by an exponential or a power-law function.
The observed shape of the Fermi bubbles seems to suggest an exponential halo.
If the velocity of the top of the envelope prevails the sound velocity in the halo, then the envelope may reach the size of $10$ kpc.
\item The surface of the top of the envelope propagates with acceleration in the halo.
As a result, Rayleigh-Taylor instabilities are developed and they will destroy the bubble envelope at the top.
We expect excitation of hydrodynamic instabilities and generation of a hydrodynamic turbulence there.
\item
Turbulent motions act as a source of waves, which are manifested as a hierarchy of eddies,
and act as a direct source of energy to the MHD waves (via the Lighthill mechanism).
For small Mach numbers, a small fraction of the power radiated by the turbulent motion is Afv\'en waves.
\item
The coefficients of the spatial and momentum diffusion of the system of nonlinear kinetic equations of the cosmic ray distribution function
are derived from the spectrum of MHD waves.
These coefficients were calculated analytically,
but we were unable to estimate the numerical values for the bubbles
because of the lack of available observation on the wave spectrum.
\item
We estimated roughly the spatial and momentum diffusions of cosmic rays in the envelope from the data of
gamma-ray and microwave radiations from the Fermi bubbles.
\item
We concluded that the observed gamma-ray and microwave radiations from the envelope of the Fermi Bubbles are generated by cosmic ray electrons only.
Contribution of cosmic ray protons can be neglected.
\item
We prefer the model that GeV cosmic ray electrons from supernova remnants in the Galactic disk
are re-accelerated in situ in the bubbles to TeV energy range.
With the help of a divergent flow, this can reproduce the data of both gamma-ray and microwave observations.
\item
On the other hand, high energy cosmic ray protons can escape the bubbles and reach the Earth.
Cosmic ray protons from supernova remnants can be accelerated in the bubbles by supersonic turbulence to higher energies.
We found that the escape high energy protons that arrive the Earth can reproduce the spectrum and flux of cosmic rays in the range
$10^{15}\sim 10^{18}$ eV (from the `knee' to the `ankle') observed near Earth.
\end{itemize}

\noindent
{\it Acknowledgements}

First of all, let us thank colleagues who participated as co-authors of publications, mentioned in the list of references.
It was a great joy for us to collaborate with them, and a significant contribution for this review we got from their collaboration.

All of them we thank very much and keep memory about any of you.

We are grateful for the nice atmosphere where we spent time participating in seminars and private talks with our colleagues from:
P. N. Lebedev Institute of Physics (Russia),
National Central University (Taiwan),
The University of Hong Kong (Hong Kong),
Max-Planck-Institut f\"ur Extraterrestrische Physik (Germany),
Institute of Space and Astronautical Science (Japan),
University of Bristol (UK), etc.

We are grateful to the International Space Science Institute (ISSI(Bern) and ISSI-BJ (Beijing)) which organized several workshops,
whose informal and warm atmosphere helped us to understand some problems on Fermi Bubbles origin.

{\color{green}
}

\end{document}